\documentclass[a4paper,11pt]{article}

\pdfoutput=1

\usepackage{jcappub}
\usepackage{amsmath}
\usepackage{amsfonts}
\usepackage{amssymb}
\usepackage{mathtools}
\usepackage{graphics}
\usepackage{citesort}
\usepackage{graphicx}
 \usepackage{url}
 \usepackage{xcolor}
\usepackage{soul}
\usepackage[utf8]{inputenc}

\usepackage{hyperref}

\title{Dispelling a myth on dense neutrino media: fast pairwise conversions depend on energy}

\author{Shashank Shalgar}
\author{and Irene Tamborra}
\affiliation{Niels Bohr International Academy and DARK, Niels Bohr Institute, University of Copenhagen, Blegdamsvej 17, 2100, Copenhagen, Denmark}

\emailAdd{shashank.shalgar@nbi.ku.dk}
\emailAdd{tamborra@nbi.ku.dk}

\abstract{
Pairwise conversions of neutrinos have complicated the challenging goal of quantifying the relevance of neutrino microphysics in compact astrophysical objects, limiting the ability to perform numerical simulations and encouraging the employment of semi-analytical tools, such as the linear stability analysis. Given the high neutrino density, the dependence of pairwise conversions on the neutrino energy has been deemed to play a negligible role. We show that  fast pairwise conversions are  affected by the neutrino energy, especially in the non-linear regime. An earlier onset of flavor conversions and a higher oscillation frequency are found as the vacuum frequency increases (i.e., the neutrino energy decreases); the oscillation periodicity, otherwise present, is gradually destroyed as the vacuum frequency increases. Such effects are, however, not further exacerbated by the inclusion of spectral energy distributions for neutrinos, contrary to what was found for  ``slow'' neutrino-neutrino conversions. In addition, our findings highlight the dependence of fast pairwise conversions  on the neutrino mass ordering for realistic values of the neutrino vacuum frequency.  Our findings highlight the limitations intrinsic to widely adopted approximations and extrapolations based on the linear regime of fast conversions. In order to gauge their possible impact on the source physics,  a more sophisticated modeling of pairwise conversions is necessary.

}

\begin{document}
\maketitle
\flushbottom

\section{Introduction}
\label{sec:intro}

Neutrino flavor conversions should play a significant, but yet poorly understood, role in dense astrophysical objects, such as core-collapse supernovae, compact binary mergers, as well as in the early Universe~\cite{Mirizzi:2015eza,Horiuchi:2017sku,Chakraborty:2016yeg,Duan:2010bg,Wu:2017qpc,Johns:2016enc}.
 Matter-enhanced flavor conversions are generated by the forward scattering of neutrinos on the background matter~\cite{Mikheev:1986gs, Wolfenstein:1977ue}. In addition, the flavor evolution is also affected by the coherent forward scattering of neutrinos among themselves~\cite{Chakraborty:2016yeg,Sigl:1992fn,Hannestad:2006nj,Fogli:2007bk,Duan:2005cp,Duan:2006jv, Duan:2006an}. Neutrino self-interactions  induce non-linear effects on the flavor evolution, leading to a rich phenomenology, still vastly unexplored. A peculiar feature of neutrino-neutrino interactions is that they strongly depend on the angle between the momenta of the interacting neutrinos. 

The impact of neutrino--neutrino interactions on the flavor evolution is challenging to predict because of the non-linear nature of the interaction. Therefore, numerical implementations have been adopted to investigate the flavor conversion phenomenology, building on the first attempts proposed in Refs.~\cite{Fogli:2007bk,Duan:2006an} and involving different levels of approximations. In addition, semi-analytical techniques, such as the linear stability analysis~\cite{Banerjee:2011fj}, have been massively employed and revealed to be especially useful in the cases most difficult to tackle numerically. 

Recently, the possibility that pairwise  conversions of neutrinos occur when the neutrino density is high has been raised~\cite{Sawyer:2015dsa,Sawyer:2005jk,Sawyer:2008zs,Chakraborty:2016lct,Dasgupta:2016dbv,Izaguirre:2016gsx}.
In this case, ``fast'' pairwise conversions would develop at a rate  much higher, $\sqrt{2} G_F (n_{\nu}-n_{\bar\nu}) \simeq$ few m$^{-1}$ (with  $n_{\nu, (\bar\nu)}$ the (anti)neutrino density and $G_F$ the Fermi constant), than the typical rate of neutrino oscillations in vacuum, $\Delta m^2/2E \simeq 0.1$~km$^{-1}$ (where $\Delta m^2$ is the atmospheric mass difference and $E \simeq \mathcal{O}(10)$~MeV is the typical neutrino energy).  Fast pairwise conversions have received a lot of attention lately as they could  potentially lead to flavor decoherence, e.g.~in the proximity of the neutrino decoupling region, with major implications on the source physics~\cite{Tamborra:2017ubu,Shalgar:2019kzy,Abbar:2019zoq, Abbar:2018shq,DelfanAzari:2019tez,Nagakura:2019sig,Morinaga:2019wsv,Wu:2017drk,Wu:2017qpc,Xiong:2020ntn,Bhattacharyya:2020dhu}.

Unlike the ``slow''  conversions due to neutrino-neutrino interactions, fast pairwise conversions have been postulated to be exclusively triggered by the neutrino-neutrino potential even in the absence of the vacuum frequency, if seed perturbations exist~\cite{Sawyer:2015dsa,Sawyer:2005jk,Sawyer:2008zs,Chakraborty:2016lct,Dasgupta:2016dbv,Izaguirre:2016gsx}. For this reason, the vacuum oscillation frequency has been neglected in most of the literature on fast pairwise conversions, see e.g.~Ref.~\cite{Chakraborty:2016lct} for a discussion on the matter.

Fast pairwise conversions have been deemed to be driven by the angular distributions of neutrinos and antineutrinos. One of the necessary conditions leading to the development of fast pairwise conversions is the existence of electron neutrino lepton number (ELN) crossings~\cite{Izaguirre:2016gsx,Abbar:2017pkh}. The existence of ELN crossings means that, in order for fast neutrino conversions to occur at a given point, the number of electron neutrinos has to be larger than the one of electron antineutrinos in some direction and smaller in another direction. 
To explore whether favorable conditions for fast pairwise conversions exist, several methods have been proposed to search for ELN crossings in hydrodynamical simulations~\cite{Abbar:2020fcl,Dasgupta:2018ulw,Johns:2019izj}.

Due to the numerical complexity of the problem, most of the existing work on fast  pairwise conversions is based on the linear stability analysis~\cite{Izaguirre:2016gsx,Martin:2019gxb,Yi:2019hrp,Capozzi:2017gqd,Abbar:2019zoq, Abbar:2018shq,DelfanAzari:2019tez,Nagakura:2019sig,Morinaga:2019wsv}. 
Although very helpful,  the oversimplifications due to the computational challenges and the wide employment of the linear stability analysis have led to the obfuscation of some relevant features of fast pairwise conversions.  For example, recent work highlighted the dynamical role played by neutrino advection in hindering the growth of flavor instabilities~\cite{Shalgar:2019qwg}, an effect that cannot be reproduced by means of the stability analysis.

In this paper, we attempt a direct comparison between the linear stability analysis and the numerical solution of the neutrino flavor evolution within a simplified setup and dispel some myths surrounding fast pairwise conversions. In particular, although the occurrence of ELN crossings is expected in the proximity of the decoupling region where the neutrino-neutrino potential is much larger than the vacuum frequency~\cite{Shalgar:2019kzy, Johns:2019izj,Tamborra:2017ubu,Izaguirre:2016gsx},  the vacuum mixing parameters cannot be ignored, and they affect the non-linear regime of fast conversions.

Our work focuses on the linear and non-linear regimes of pairwise conversions of neutrinos and aims to highlight the modifications induced in the flavor conversion phenomenology when the neutrino energy dependence is taken into account. Notably, a first attempt to explore the physics of fast flavor conversions in the non-linear regime was presented in  Ref.~\cite{Dasgupta:2017oko} in the context of a four beam neutrino model and in the limit of vanishing vacuum frequency; the same limit was also explored in Ref.~\cite{Johns:2019izj}. As a consequence of the vanishing vacuum frequency, Refs.~\cite{Dasgupta:2017oko,Johns:2019izj} reported a periodic trend in the fast flavor conversion probability  in the non-linear regime. We find that the periodic trend is destroyed for realistic values of the vacuum frequency; moreover, 
 simple formal analogies widely adopted to intuitively explain the $\nu$--$\nu$ interaction physics, such as the pendulum one proposed in Ref.~\cite{Hannestad:2006nj}, are not applicable to the case of fast pairwise conversions, as also pointed out in~\cite{Johns:2019izj,Bhattacharyya:2020dhu}. Intriguingly, our findings highlight that fast pairwise conversions depend on the mass ordering, contrarily to the conclusions drawn in Ref.~\cite{Dasgupta:2017oko} by relying on a vanishing vacuum frequency. Given the highly non-linear nature of the system studied here, we refrain from providing an analytical understanding of our findings and mostly focus on discussing insightful examples of how the conversion physics is modified.

This paper is organized as follows. 
We briefly review the neutrino equations of motion and the physics of fast pairwise conversions in Sec.~\ref{sec2}. The dependence of the linear stability analysis on the vacuum frequency is discussed in  Sec.~\ref{sec3}. In Sec.~\ref{sec4} we focus on the non-linear regime and explore the dependence of fast neutrino conversions on the vacuum mixing parameter and on the mass ordering. Our results are extended from a single-energy to a multi-energy framework in Sec.~\ref{sec5}. Section~\ref{sec6} focuses on how our findings are further affected when the reflection symmetry is broken. Finally, conclusions are reported in Sec.~\ref{sec7}.

\section{Neutrino equations of motion}
\label{sec2}

The  equations of motion describing the neutrino flavor evolution can be written using the Wigner transformed density matrices to encode the flavor of neutrinos. For simplicity, we work in the two flavor approximation $(\nu_e,\nu_x)$ and thus we have two density matrices $\rho$ and $\bar{\rho}$ for neutrinos and antineutrinos of momentum $\vec{p}$, respectively. Each one is a $2 \times 2$ matrix and evolves in accordance with the Heisenberg equation of motion:
\begin{eqnarray}
        i\frac{d\rho(\vec{p})}{dt} &=& \left[H(\vec{p}),\rho(\vec{p})\right] \\
        i\frac{d\bar{\rho}(\vec{p})}{dt} &=& \left[\bar{H}(\vec{p}),\bar{\rho}(\vec{p})\right]\ .
        \label{eom}
\end{eqnarray}
The Hamiltonian, that governs the temporal evolution of the density matrix, has three contributions: the vacuum term, the matter term, and the neutrino-neutrino term:
\begin{eqnarray}
        H(\vec{p}) &=& H_{\textrm{vac}}+H_{\textrm{mat}}+H_{\nu\nu}\ ,\\
        \bar{H}(\vec{p}) &=& -H_{\textrm{vac}}+H_{\textrm{mat}}+H_{\nu\nu}\ .
\end{eqnarray}
In the two flavor approximation, the three contributions of the Hamiltonian are  defined as follows
\begin{eqnarray}
        H_{\textrm{vac}} &=& 
\frac{\omega}{2}\begin{pmatrix}
        -\cos 2 \theta_{\textrm{V}} & \sin 2 \theta_{\textrm{V}} \\
        \sin 2 \theta_{\textrm{V}} & \cos 2 \theta_{\textrm{V}}
\end{pmatrix},
\\
        H_{\textrm{mat}} &=& 
        \begin{pmatrix}
                \sqrt{2} G_{\textrm{F}} n_{e} & 0 \\
                0 & 0
        \end{pmatrix},
\\
        H_{\nu\nu} &=& \mu \int d\vec{p^{\prime}}[\rho(\vec{p})-\bar{\rho}(\vec{p})]
        \left(1-\frac{\vec{p}\cdot\vec{p^{\prime}}}{|\vec{p}||\vec{p^{\prime}}|}\right),
\end{eqnarray}
where $\omega =  \Delta m^2/2E$ is the vacuum frequency (with the squared mass splitting $\Delta m^2 >0$ for normal mass ordering and $\Delta m^2 <0$  for inverted ordering and $E$ being the neutrino energy), $\theta_{\textrm{V}}$ is the vacuum mixing angle, $G_{\textrm{F}}$ is the Fermi constant, $n_{e}$ is the effective number density of electrons in the medium, and $\mu = \sqrt{2} G_F n_{\nu}$ is the strength of the self-interaction potential which is proportional to the neutrino number density $n_{\nu}$. 
 The first two terms of the Hamiltonian, $H_{\textrm{vac}}$ and $H_{\textrm{mat}}$, are linear in nature. On the other hand, the third term, $H_{\nu\nu}$,  is dependent on the density matrices and thus makes  the equations of motion non-linear.

Throughout the paper, we assume $\mu=10^{5}$ km$^{-1}$ and $\theta_V = 10^{-6}$ to mimic the effective  suppression of mixing due to matter~\cite{EstebanPretel:2008ni}, and $n_e = 0$. Moreover, in spherical coordinates (with $\theta$ being the polar angle and $\phi$ the azimuthal one) and
under the assumption of azimuthal symmetry, the neutrino-neutrino term of the Hamiltonian becomes
\begin{eqnarray}
     H_{\nu\nu}(\theta) = 2 \pi \mu \int_{0}^{\pi} d\theta^{\prime} \sin \theta^{\prime} [\rho(\theta^{\prime})-\bar{\rho}(\theta^{\prime})](1-\cos\theta\cos\theta^{\prime})\ ;
        \label{Hnunu:sym}
\end{eqnarray}
the angles $\theta$ and $\theta^{\prime}$ are the polar angular coordinates of neutrinos, and the integration over the azimuthal angle gives a $2\pi$ factor. Equation~\ref{Hnunu:sym} is formally identical to the one used in, e.g.,~\cite{Duan:2006an}; however, in this paper, we are interested in exploring the flavor conversion physics in a simple toy model and, consequently, we do not connect  $\theta$ and $\theta^\prime$ to the geometry of the emission surface of an astrophysical source. The angular coordinates adopted  in this paper should  be interpreted as the local ones in the vicinity of the region where we solve the equations of motion.

Neutrino-neutrino interactions are a consequence of the coherent forward scattering of neutrinos with other neutrinos: the incident neutrino exchanges momentum with the target neutrino. When the neutrino number density is large enough, it has been postulated that significant neutrino flavor evolution may occur as it is possible for a pair of neutrinos to coherently scatter of each other and change the flavor~\cite{Sawyer:2015dsa,Sawyer:2008zs,Sawyer:2005jk,Chakraborty:2016lct}:
\begin{eqnarray}
        \nu_{\alpha} \bar{\nu}_{\alpha} \rightarrow \nu_{\beta} \bar{\nu}_{\beta}\ ,
\end{eqnarray}
where $\alpha$ and $\beta$ denote the neutrino flavor. Such pairwise conversions have been found to occur coherently when there is an ELN crossing~\cite{Izaguirre:2016gsx}. 

The equations of motion that govern the flavor evolution in the presence of fast pairwise conversions of neutrinos are identical to the ones describing ``slow'' neutrino self-interactions~\cite{Sigl:1992fn}. However, whereas the slow neutrino self-interactions effectively occur when the neutrino self-interaction potential is roughly comparable to the vacuum frequency, fast pairwise conversions can lead to significant flavor evolution for arbitrarily large neutrino-neutrino potential.

A necessary condition for significant flavor evolution is that the off-diagonal component of the Hamiltonian is not very small in comparison with the diagonal elements. This can be either due to a large off-diagonal term in the linear part of the Hamiltonian or  to a dynamical increase in the off-diagonal term of $H_{\nu\nu}$ caused by the non-linear nature of the evolution. A dynamical rise in the off-diagonal term of the density matrix is  referred to as ``flavor instability.'' 

In the limit where we can ignore the momentum-changing scatterings, the occurrence of a flavor instability has been considered to be solely determined by the (anti)neutrino angular distributions,  i.e., their relative shape and normalization~\cite{Izaguirre:2016gsx,Abbar:2017pkh}. In the case of core-collapse supernovae, conditions favoring the development of  ELN crossings have been found in the proximity of the neutrino decoupling region where neutrinos gradually transition from being trapped to the free-streaming regime and in the pre-shock region~\cite{Tamborra:2017ubu,Shalgar:2019kzy,Abbar:2019zoq, Abbar:2018shq,DelfanAzari:2019tez,Nagakura:2019sig,Morinaga:2019wsv,Wu:2017drk,Wu:2017qpc,Xiong:2020ntn}. It should be noted that the cross-section of $\nu_e$ with nucleons is larger than the one of $\bar\nu_e$ with nucleons, resulting in a larger decoupling radius for $\nu_e$ in comparison to $\bar\nu_e$, see e.g.~\cite{Tamborra:2017ubu,Shalgar:2019kzy}; thus, there is a region near the decoupling region where $\nu_e$'s are more isotropically distributed than $\bar\nu_e$'s whose distribution is slightly forward peaked. This  naturally provides a benchmark scenario where ELN crossings can occur and is a convenient case of study for our purposes.

\section{Energy dependent linear stability analysis}
\label{sec3}

In this section, we introduce the linear stability analysis tool commonly adopted to gauge whether favorable conditions for fast pairwise conversions exist.  However, the linear stability analysis for exploring the existence of flavor instabilities due to pairwise conversions has been derived for vanishing vacuum frequency~\cite{Chakraborty:2016yeg,Izaguirre:2016gsx}. This simplification was naively justified because $\mu \gg \Delta m^2/2E$. Since our main goal is to investigate the dependence of fast pairwise conversions on $\omega =\Delta m^2/2E$, we now include the vacuum term of the Hamiltonian, see also Refs.~\cite{Airen:2018nvp,Chakraborty:2016yeg}.

\subsection{Growth rate of the flavor instability}
For the sake of simplicity, we consider an ensemble of neutrinos such that they are approximately in the electron flavor eigenstates and assume that the off-diagonal components are much smaller than the diagonal components. Hence, we can write the density matrices in the following form:
\begin{eqnarray}
        \rho(\vec{p}) = 
        \begin{pmatrix}
                \rho_{ee}(\vec{p}) & \rho_{ex}(\vec{p}) \\
                \rho_{ex}^{*}(\vec{p}) & 0
        \end{pmatrix}
\quad \quad
        \bar{\rho}(\vec{p}) =
        \begin{pmatrix}
                \bar{\rho}_{ee}(\vec{p}) & \bar{\rho}_{ex}(\vec{p}) \\
                \bar{\rho}_{ex}^{*}(\vec{p}) & 0
        \end{pmatrix}\ ,
\end{eqnarray}
with $\rho_{ex} \ll \rho_{ee}$ and $\bar{\rho}_{ex} \ll \bar{\rho}_{ee}$. If the flavor evolution is collective in nature, all momentum modes evolve with the same frequency denoted by $\Omega$,
\begin{eqnarray}
        \rho_{ex}(\vec{p}) &=& Q_{\vec{p}} e^{-i\Omega t} \nonumber\\
        \bar{\rho}_{ex}(\vec{p}) &=& \bar{Q}_{\vec{p}} e^{-i\Omega t}\ .
\label{ansatz}
\end{eqnarray}
Plugging this ansatz in Eq.~\ref{eom},  we get the following characteristic equations:
\begin{eqnarray}
\Omega Q_{\theta} = H_{ee} Q_{\theta} - 2 \pi \rho_{ee} \int\left(Q_{\theta^{\prime}} - \bar{Q}_{\theta^{\prime}}\right)
\times (1-\cos\theta\cos\theta^{\prime}) \sin\theta^{\prime} d\theta^{\prime}\nonumber\\
\Omega \bar{Q}_{\theta} = \bar{H}_{ee} \bar{Q}_{\theta} - 2 \pi \bar{\rho}_{ee} \int\left(Q_{\theta^{\prime}} - \bar{Q}_{\theta^{\prime}}\right)
\times (1-\cos\theta\cos\theta^{\prime}) \sin\theta^{\prime} d\theta^{\prime}\ ,
\label{lineom}
\end{eqnarray}
where we adopt $Q_\theta$ instead of $Q_{\vec{p}}$ in the single-energy approximation. Moreover, the term $H_{ee}$ includes the vacuum term and it therefore depends on the neutrino energy $E$.
An inspection of Eq.~\ref{lineom} implies that $Q_{\theta}$ and $\bar{Q}_{\theta}$ are
\begin{eqnarray}
Q_{\theta} \propto a + c \cos\theta\ \mathrm{and}\ 
\bar{Q}_{\theta} \propto a + c \cos\theta\ ,
\label{Qform}
\end{eqnarray}
with $a$ and $c$ being undetermined constants depending on the neutrino energy.

By substituting Eq.~\ref{Qform} in Eq.~\ref{lineom}, we obtain 
\begin{eqnarray}
        \begin{vmatrix}
        I[1]-1 & -I[\cos\theta] \\
        I[\cos\theta] & -I[\cos^{2}\theta]-1
        \end{vmatrix} = 0\ ,
\label{det}
\end{eqnarray}
where
\begin{equation}
        I[f(\theta)] = 2\pi \int_{0}^{\pi} d\theta \sin(\theta) f(\theta) \left[\frac{\rho_{ee}(\theta)}{2\omega+A-B \cos\theta} - 
        \frac{\bar{\rho}_{ee}(\theta)}{-2\omega+A-B \cos\theta}\right]\ ,
        \end{equation}
        and
        \begin{eqnarray}
        A = \Omega - 2\pi \int_{0}^{\pi} d\theta \sin(\theta) (\rho_{ee}-\bar{\rho}_{ee}) \ \mathrm{and}\ 
        B =  - 2\pi \int_{0}^{\pi} d\theta \sin\theta (\rho_{ee}-\bar{\rho}_{ee}) \cos\theta\ . 
\end{eqnarray}
It is possible to have multiple values of $\Omega$ for which Eq.~\ref{det} is satisfied, but only the solution with $\textrm{Im}(\Omega) > 0$ implies a flavor instability.

\subsection{Results of the linear stability analysis}
In order to investigate the dependence of the flavor instability  on the initial angular distribution of $\nu_e$ and $\bar\nu_e$, we assume the following distributions:
\begin{eqnarray}
\label{spec1} 
        \rho_{ee}(\theta) &=& {\mathrm{const.}}\\ 
        \bar{\rho}_{ee}(\theta) &=& (1-y_{1})+2 y_{1} \exp\left(-\frac{\theta^{2}}{2}\right) \quad \textrm{for } 0 < \theta < \pi\ ,
\end{eqnarray}
where the $\nu_e$ density matrix is normalized such that
        $\int_{0}^{\pi} \int_{0}^{2\pi} \rho_{ee}(\theta)  d\phi \sin\theta d\theta = 4\pi$.
The growth rate (see Eq.~\ref{ansatz}), Im($\Omega$), increases as a function of $y_{1}$. The latter is assumed to be $y_1 =0.05$, we also consider one energy mode for neutrinos and antineutrinos with $E = 10$~MeV.
\begin{figure}
        \begin{center}
        \includegraphics[width=0.49\textwidth]{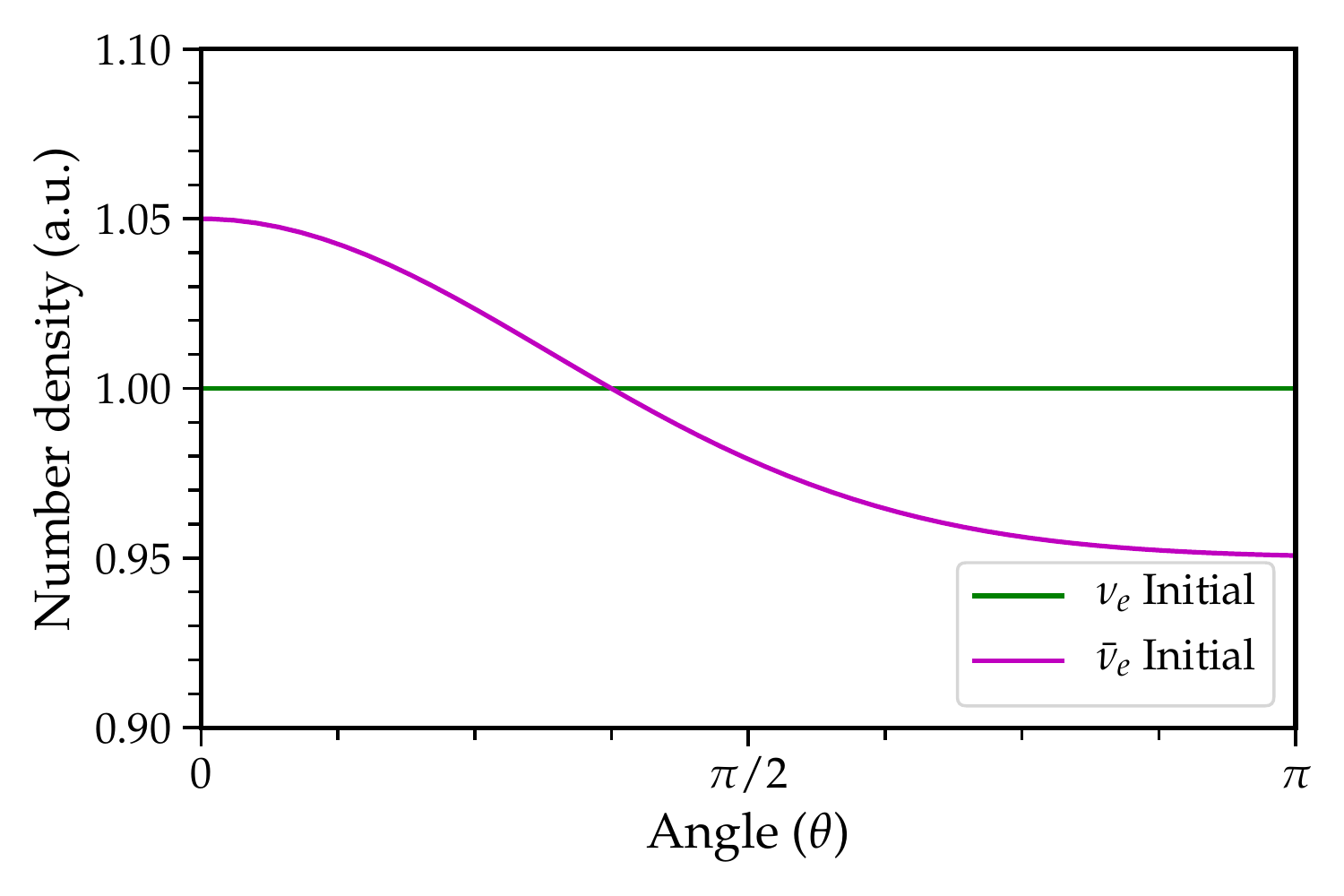}  
        \includegraphics[width=0.49\textwidth]{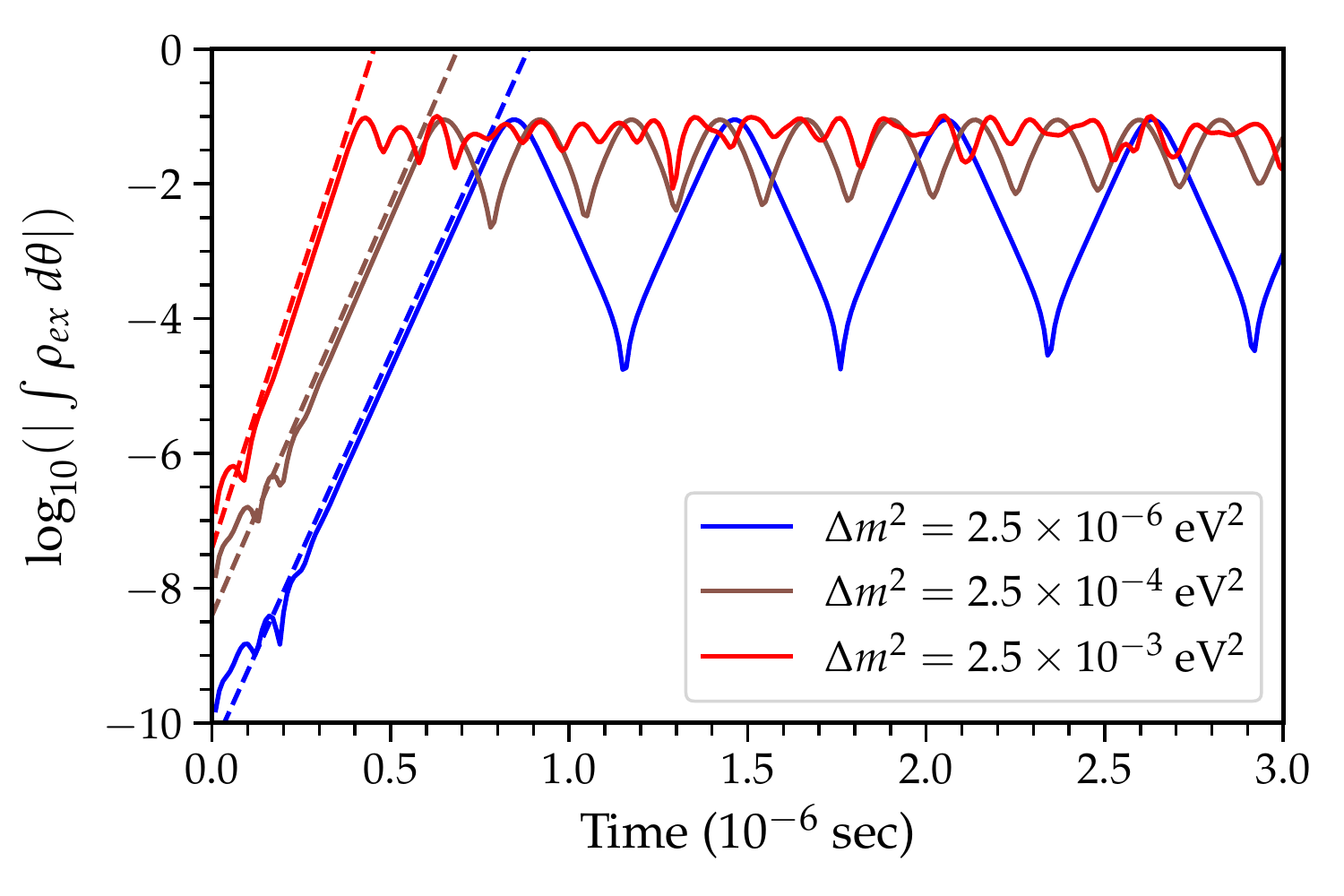}
        \end{center}
        \caption{{\it Left:} Initial angular distributions for $\nu_{e}$ (in green) and $\bar{\nu}_{e}$ (in magenta) as  functions of $\theta$ and for $y_1=0.05$, see Eq.~\ref{spec1}. 
{\it Right:} Growth rate  of the angle-integrated off-diagonal term of the  neutrino density matrix (dashed lines) as a function of time for the initial flavor configuration introduced in Eq.~\ref{spec1}. The solid lines show the evolution of the  off-diagonal term of the neutrino density matrix calculated using numerical simulations, while the growth rate is calculated using the linear stability analysis. The blue, brown, and red lines show the evolution of the angle-integrated off-diagonal term of the density matrix for $\Delta m^{2} = 2.5 \times 10^{-6}, 2.5 \times 10^{-4}$, and $2.5 \times 10^{-3} \textrm{ eV}^{2}$, respectively. The onset of the non-linear regime and the oscillation frequency and amplitude depend on $\Delta m^{2}$.
        }
        \label{Fig1}
\end{figure}

The angular distributions of $\nu_e$'s and $\bar\nu_e$'s introduced above are plotted as functions of $\theta$ in the left panel of Fig.~\ref{Fig1} and they exhibit one ELN crossing.  For this initial configuration, the linear stability analysis predicts a non-zero growth rate, as shown in the right panel of Fig.~\ref{Fig1} (dashed lines) for different values of $\Delta m^2$ (note that exploring the $\omega$-dependence of fast pairwise conversions by varying $\Delta m^2$ is equivalent to change $E$ for our purposes). The growth rates of the off-diagonal term of the density matrix for different values of  $\Delta m^{2}$ do not drastically differ among each other in the linear regime; however,  as we will discuss in Sec.~\ref{sec4}, the evolution of the off-diagonal term of the neutrino density matrix shows a marked  dependence on $\Delta m^{2}$ in the non-linear regime.

\section{Impact of the vacuum mixing parameter: single-energy configuration}
\label{sec4}

In order to explore the effect of the vacuum frequency, $\omega = \Delta m^2/2E$, on the evolution of the neutrino flavor due to fast pairwise conversions, we now focus on the numerical computation of the flavor evolution for a neutrino ensemble with one energy mode for neutrinos and antineutrinos ($E=10$~MeV). We assume the angular distributions introduced in Eq.~\ref{spec1}.  First, we focus on exploring the flavor evolution as a function of $\Delta m^2$ for $\Delta m^2 > 0$. Then,  we discuss the dependence of the flavor evolution on the mass ordering.

\subsection{Impact of the vacuum mixing parameter}
The right panel of Fig.~\ref{Fig1} shows the growth rate of the angle-integrated off-diagonal term of the neutrino density matrix for different values of $\Delta m^2$ and its following evolution in the non-linear regime. The onset of the non-linear regime occurs earlier for larger $\Delta m^2$, in agreement with the findings of Ref.~\cite{Dasgupta:2017oko}; this is due to the fact that a larger $\Delta m^2$ induces a larger off-diagonal seed in the density matrix that leads the instability to grow faster. In addition,  quasi-periodic modulations of $\int \rho_{ex} d\theta$ appear in the non-linear regime, and the oscillation frequency is higher for larger values of $\Delta m^2$. 
\begin{figure}
        \includegraphics[width=0.49\textwidth]{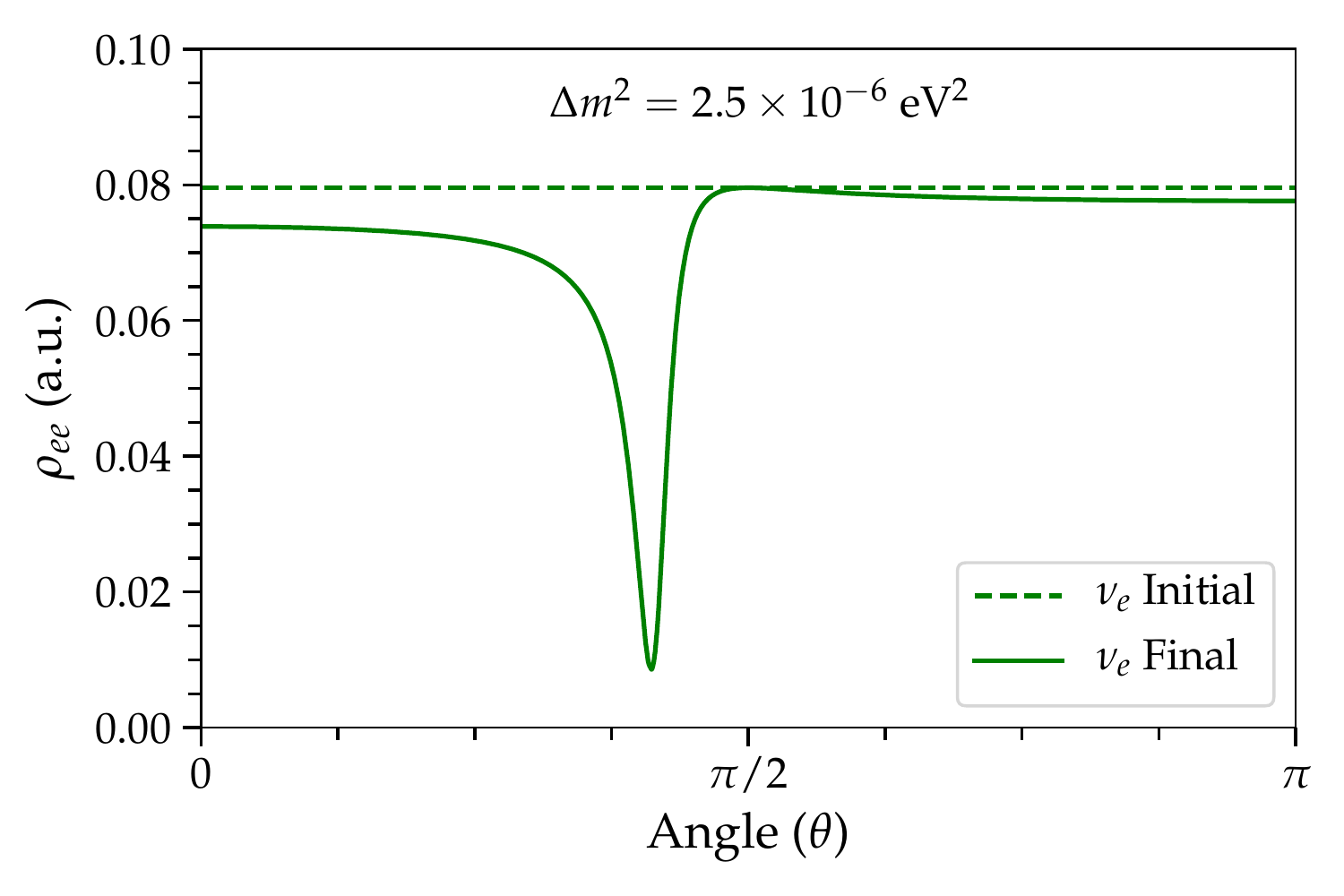}
        \includegraphics[width=0.49\textwidth]{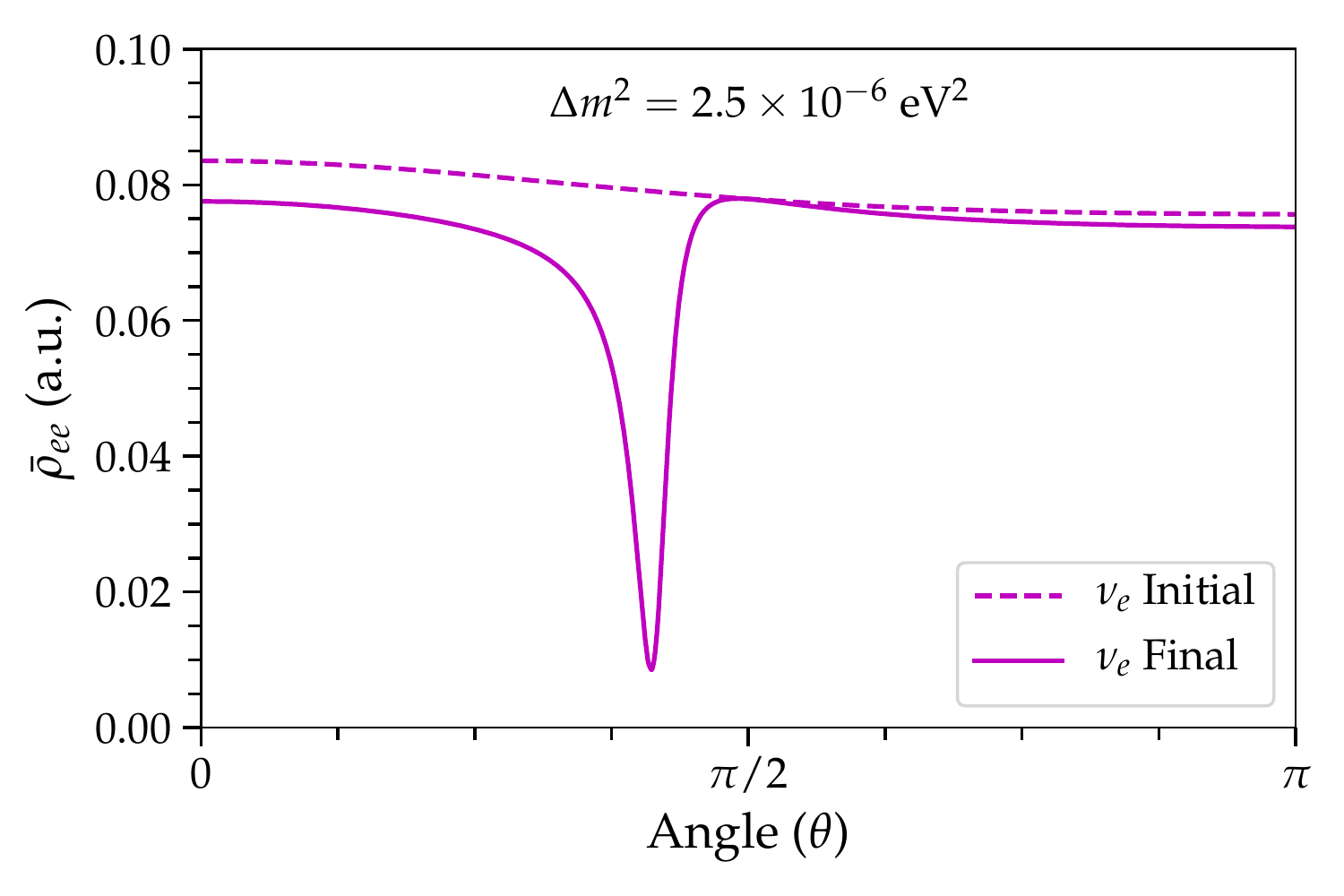}
        \includegraphics[width=0.49\textwidth]{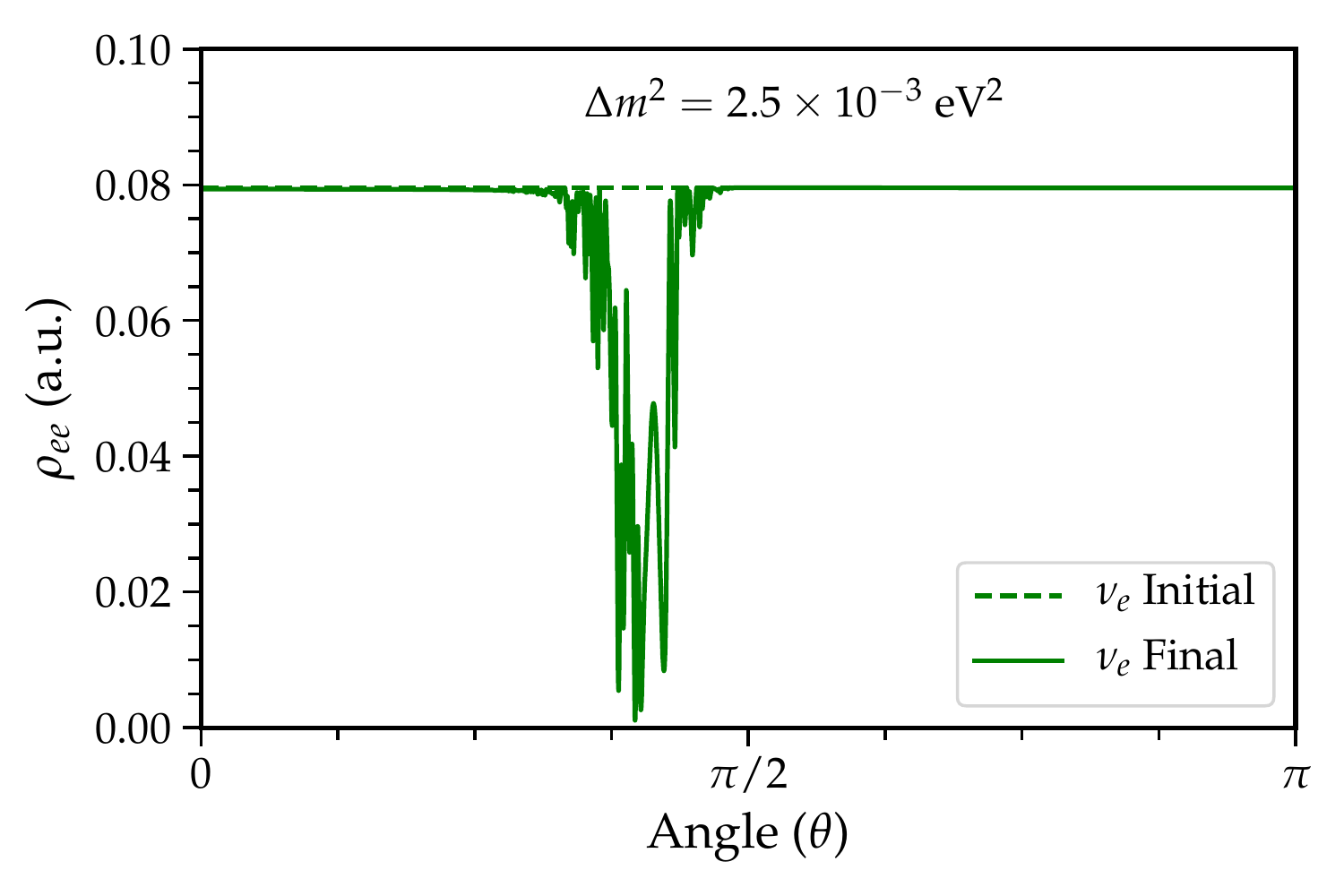}
        \includegraphics[width=0.49\textwidth]{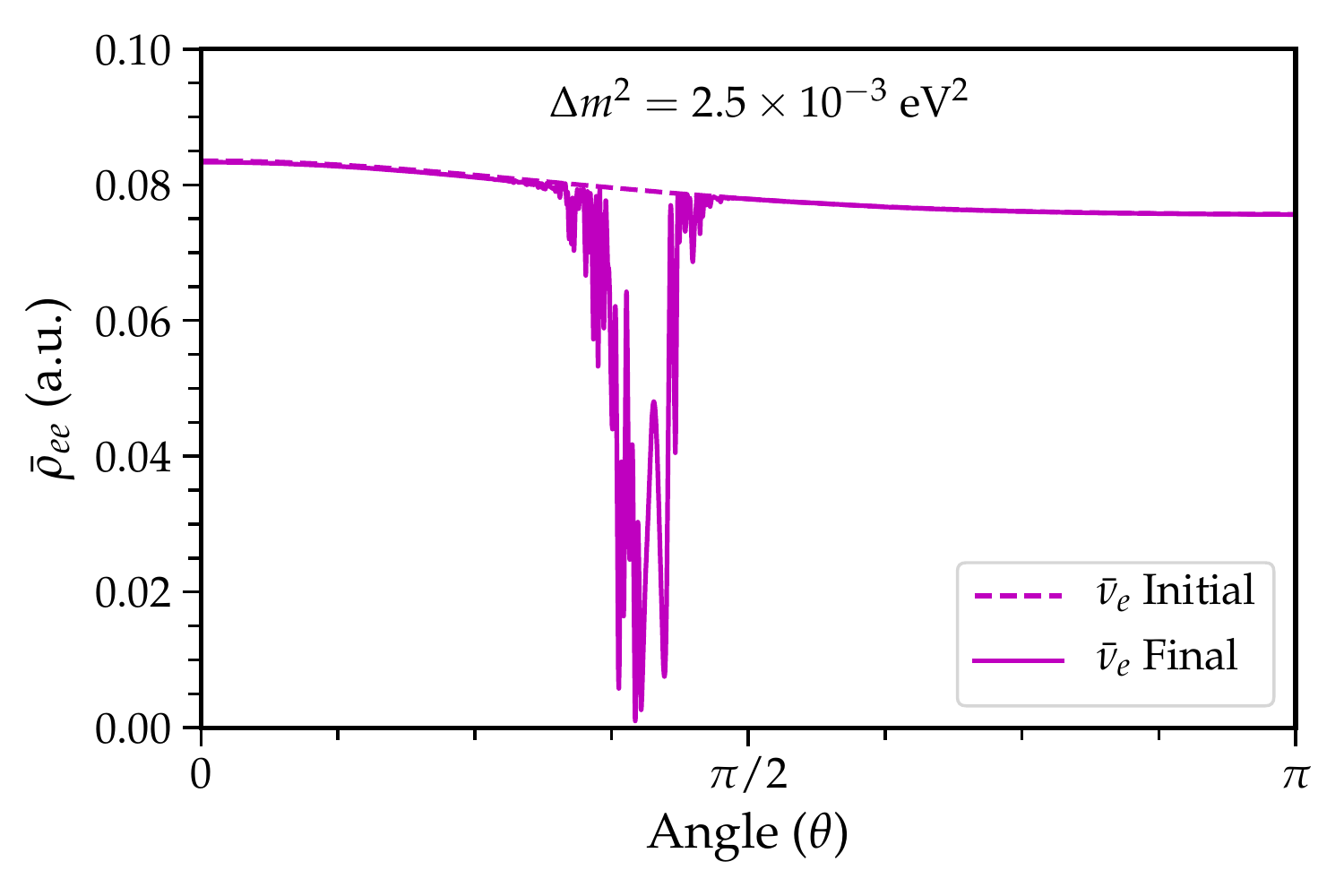}
        \caption{
Angular distributions of $\nu_{e}$ (in green, on the left) and $\bar{\nu}_{e}$ (in magenta, on the right) as a function of $\theta$
for the single-energy configuration ($E=10$ MeV). In the top panel the solid lines are the angular distributions at a representative time when the transition probability is near the peak ($t=2.63\times 10^{-6}$ sec), while the dashed lines represent the initial distributions. The top (bottom) panels  are for $\Delta m^2 = 2.5 \times 10^{-6}$~eV$^2$  ($\Delta m^{2} = 2.5 \times 10^{-3}$~eV$^{2}$)
 The evolution of the neutrino flavor is  quasi-periodic and the deviation of the final distributions from the initial ones depends on $\theta$ and $t$ significantly. }
        \label{Fig2}
\end{figure}
In order to better grasp the role of $\omega$ in fast pairwise conversions, Fig.~\ref{Fig2} shows the initial and final (at $t=3\times 10^{-6}$~sec)  angular distributions of $\nu_e$ and $\bar\nu_e$ as  functions of $\theta$. The flavor instability starts to develop in the angular bins where the ELN crossing occurs and proceeds periodically, spreading in the neighboring angular bins. This is clearly shown in the animations provided as \href{https://sid.erda.dk/share_redirect/DuI6O4k9Py/index.html}{Supplemental Material}.  However, while we find a behavior remarkably similar to the one of ``slow'' bipolar oscillations~\cite{Hannestad:2006nj}  in the limit of $\Delta m^2  \rightarrow 0$, as also discussed in Refs~\cite{Dasgupta:2017oko,Johns:2019izj}, where the only dimension-full quantity in the equations of motion is $\mu$, two different frequencies ($\Delta m^2/2E$ and $\mu$) affect the flavor evolution when $\Delta m^2 \gg 0$. 
 As seen in the bottom panels of Fig.~\ref{Fig2},   the angular distributions acquire a more complex structure for larger $\Delta m^2$ (e.g., for $\Delta m^2$ corresponding to the measured atmospheric mass difference), and the periodicity of the flavor evolution is slowly lost as shown by the red line in the right panel of Fig.~\ref{Fig1}.

The animations highlight that, as time increases,  the angular distributions keep developing fine structures. Consequently, it is extremely difficult to assess whether the angle-averaged flavor distribution ever reaches a steady state. Comparing the top and bottom panels of Fig.~\ref{Fig2}, and as also shown in the right panel of Fig.~\ref{Fig1}, it is clear that the time scale required to achieve a  steady-state configuration--if ever reached--depends on the vacuum frequency. Figures~\ref{Fig1} and \ref{Fig2} highlight that $\Delta m^2 \neq 0$ plays a fundamental role in determining the frequency and the development of flavor conversions despite being much smaller than $\mu$.

The temporal evolution of flavor can be intuitively explained by looking at the animations (see \href{https://sid.erda.dk/share_redirect/DuI6O4k9Py/index.html}{Supplemental Material}). First $\Delta m^2 \neq 0$ drives the polarization vectors to precess at a certain inclination angle with respect to the $z$ axis, as it  was found in the ``slow'' $\nu$--$\nu$ interaction case, and this triggers the growth of $\rho_{ex}$. However,  the polarization vectors are strongly attracted towards the $z$ axis again where they return in the $t=0$ configuration, and the whole process repeats with a certain periodicity.
 This trend immediately suggests that the  ``pendulum analogy''~\cite{Hannestad:2006nj} widely adopted to explain the $\nu$--$\nu$ interaction physics in the slow regime does not hold in the ``fast'' regime.

\begin{figure}
\begin{center}
\includegraphics[width=0.49\textwidth]{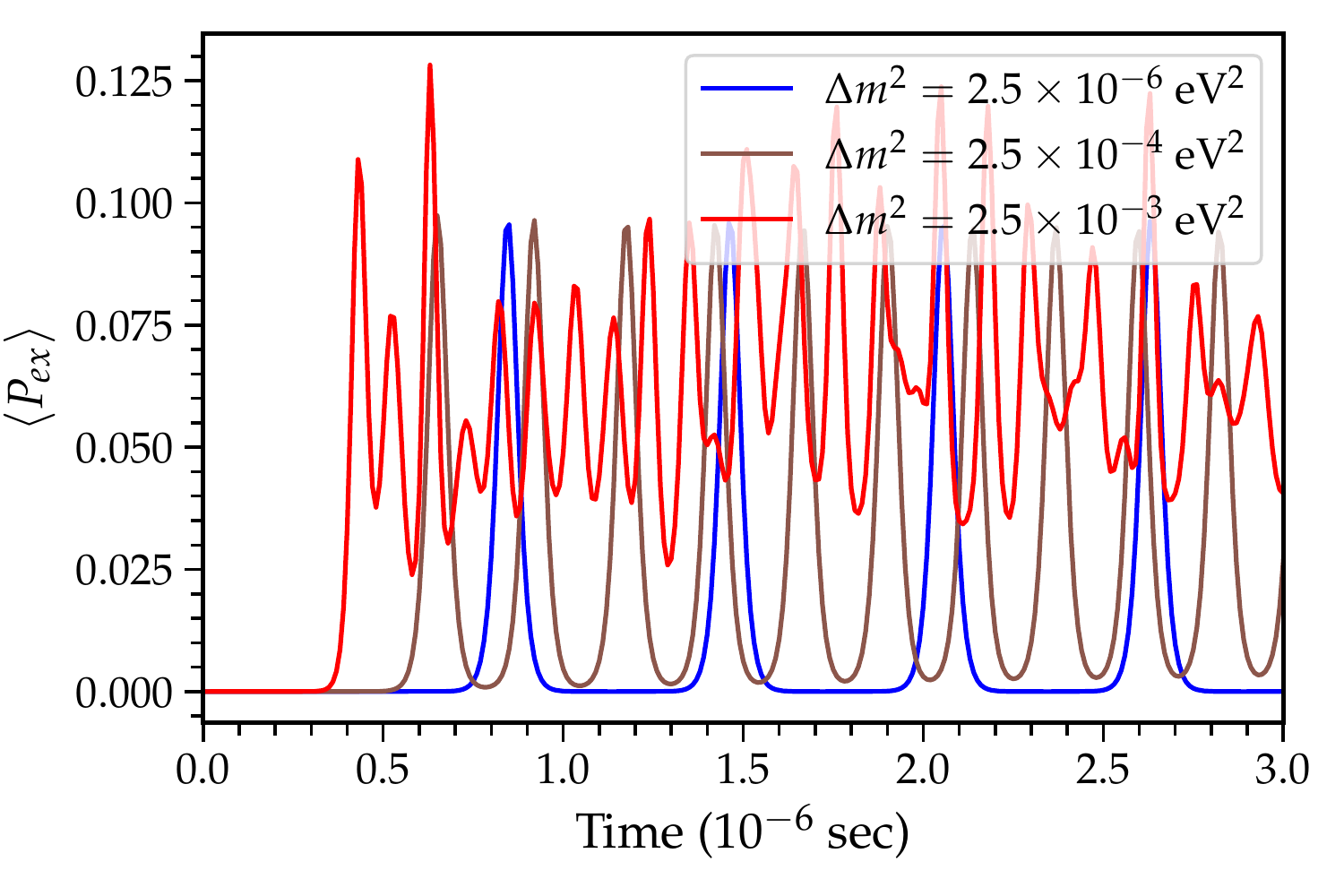}
\end{center}
\caption{Evolution of the angle-averaged transition probability $\langle P_{ex} \rangle$ as a function of time for  the single-energy case  ($E=10$~MeV) and for $\Delta m^2 = 2.5 \times 10^{-6}, 2.5 \times 10^{-4}$ and $2.5 \times 10^{-3}$ eV$^{2}$ in blue, brown, and red respectively. The onset of flavor conversions occurs earlier for larger $\Delta m^2$ and  the oscillation frequency increases as  $\Delta m^2$ increases. The  deviation from the bipolar oscillation pattern is more pronounced as  $\omega$ increases.
}
\label{Fig3}
\end{figure}

Figure~\ref{Fig3} shows the temporal evolution of the angle-averaged transition probability: 
\begin{eqnarray}
\label{eq:Pex}
\langle P_{ex} \rangle(t) =\frac{\int \rho_{xx}(\theta,t) d\theta - \int \rho_{ee}(\theta,t=0) d\theta}
{\int \rho_{xx}(\theta,t=0) d\theta - \int \rho_{ee}(\theta,t=0) d\theta}\ ,
\end{eqnarray}
for different values of $\Delta m^2$. 
A  rapid oscillatory behavior is clearly evident as typical of fast pairwise conversions. However,  the period of conversions does depend on $\Delta m^2$ as well as the oscillation amplitude. This behavior is not captured by the linear stability analysis and highlights the importance of numerical simulations in the non-linear regime. 
Notably, despite the growth rate being large as shown in the right panel of Fig.~\ref{Fig1}, the effective transition probability in Fig.~\ref{Fig3} is small. 
Figure~\ref{Fig3} highlights that, as the vacuum frequency increases, not only does the frequency of the flavor transition probability increases, but the deviation from the bipolar oscillation pattern is larger as $\omega$ increases.

Our results clearly show that, even if the $\nu$--$\nu$  potential is much larger than the vacuum frequency, the vacuum frequency cannot be ignored, and  it  affects the flavor evolution. 
 In particular, in any realistic astrophysical system, the size of the region over which neutrinos are emitted would be larger than the product of the speed of light and the time scales typically adopted in this paper; this implies that, for any hypothetical point of observation,  the time-averaged conversion probability is the physical quantity to  be considered. In light of this, Fig.~\ref{Fig3} shows that the vacuum frequency is as relevant as any other quantity in determining the flavor conversion probability. It should also be noted that, for $\Delta m^2 \rightarrow 0$, the time between two bipolar modes increases and the time-averaged flavor conversion probability tends to zero; this is not anymore the case for $\Delta m^2 \gg 0$.
The non-linear evolution of fast-pairwise flavor conversions was also explored in Ref.~\cite{Abbar:2018beu}. Despite the large ELN crossings, a modulation in the amplitude of the flavor conversions was found there, suggesting that the vacuum frequency may play a role, although this was not the focus of that paper.

\subsection{Impact of the mass ordering}

The counter-intuitive role played by the vacuum frequency on fast pairwise conversions naturally leads to investigate the impact of the neutrino mass ordering. In this section, we present the results of our numerical simulations with  $\Delta m^2 < 0$, while keeping all other parameters unchanged. 

\begin{figure}
\begin{center}
\includegraphics[width=0.49\textwidth]{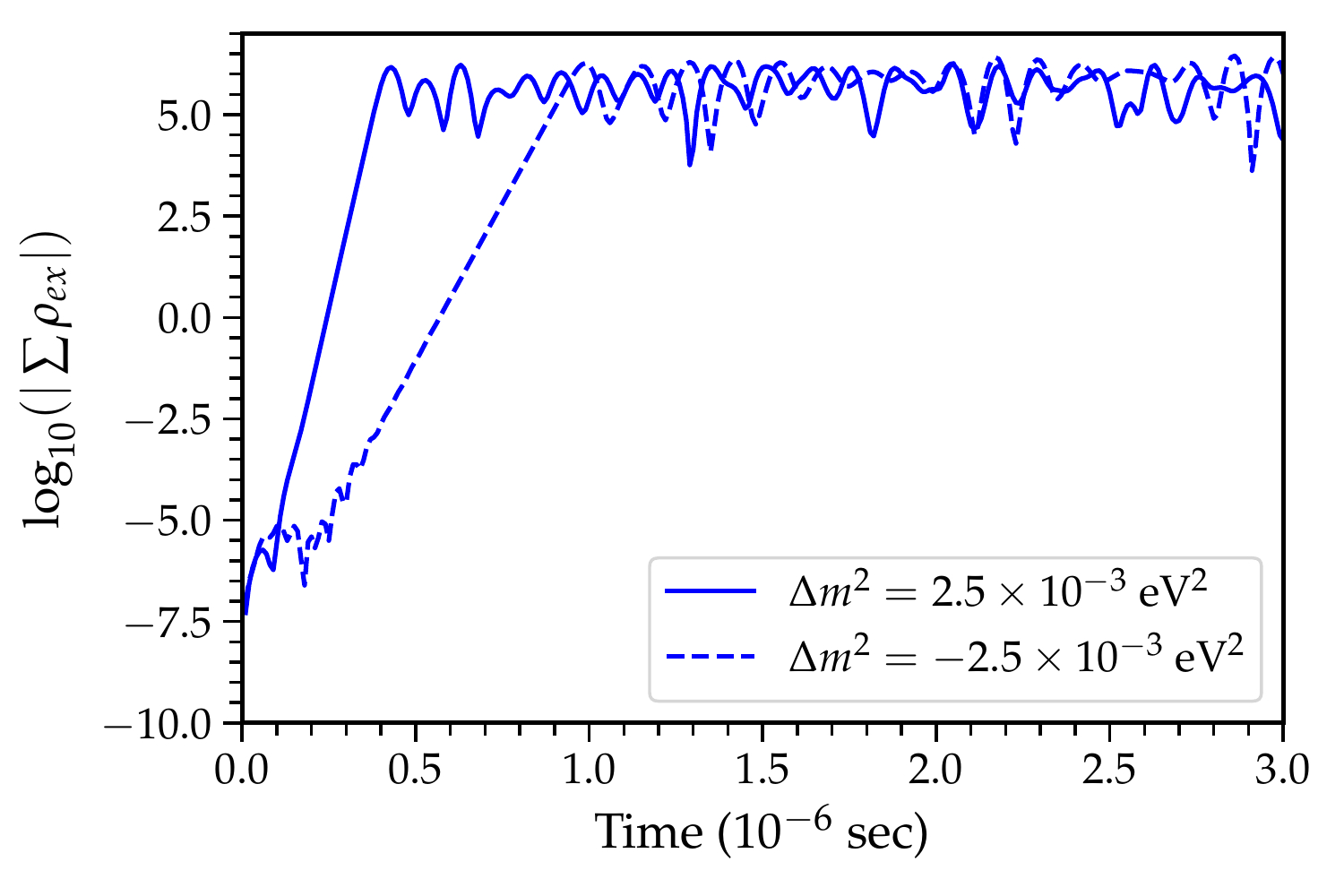}
\includegraphics[width=0.49\textwidth]{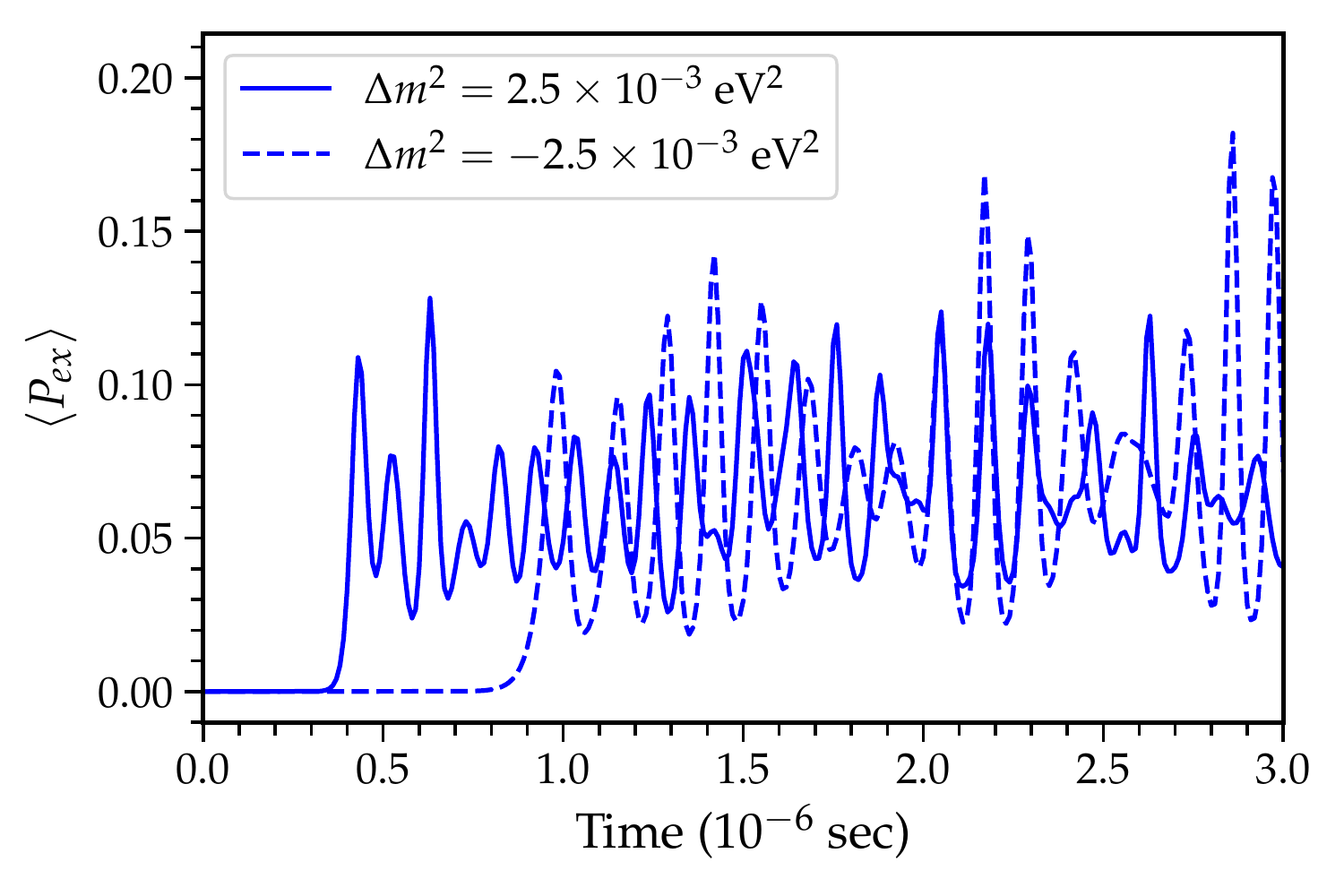}
\end{center}
\caption{{\it Left:} Evolution of the angle-integrated off-diagonal term of the  neutrino density matrix as a function of time  (see Eq.~\ref{spec1}), in normal ordering ($\Delta m^{2} = 2.5 \times 10^{-3}$ eV$^{2}$, solid curve) and inverted ordering ($\Delta m^{2} = -2.5 \times 10^{-3}$ eV$^{2}$, dashed curve). 
 {\it Right:} Evolution of the angle-averaged flavor transition probability as a function of time for $\Delta m^{2} = 2.5 \times 10^{-3}$ eV$^{2}$ (continuous curve) and $\Delta m^{2} = -2.5 \times 10^{-3}$ eV$^{2}$ (dashed curve). The onset of flavor transitions is delayed in the inverted ordering case with respect to the normal ordering one, and the oscillation amplitude is larger in inverted ordering.}
\label{Fig4}
\end{figure}

The left panel of Fig.~\ref{Fig4} shows the evolution of the off-diagonal terms of the density matrix as a function of time for normal (continuous line, $\Delta m^{2} = 2.5 \times 10^{-3}$ eV$^{2}$) and inverted (dashed line, $\Delta m^{2} = -2.5 \times 10^{-3}$ eV$^{2}$) ordering. One can see that the growth rate is steeper in normal ordering and the non-linear regime is reached earlier, as also visible in the right panel of Fig.~\ref{Fig4}. The latter shows the temporal evolution of $\langle P_{ex} \rangle$ (see Eq.~\ref{eq:Pex}) for both orderings. In the case of inverted ordering, the oscillation amplitude is larger. In fact, as one can see from the animations provided as \href{https://sid.erda.dk/share_redirect/DuI6O4k9Py/index.html}{Supplemental Material}, the system is more unstable in inverted ordering and it departs from its initial configuration a bit more at each precession, than in the normal ordering case. 
Our findings also suggest that the time required to reach the steady state configuration depends on the mass ordering (see \href{https://sid.erda.dk/share_redirect/DuI6O4k9Py/index.html}{Supplemental Material}).

Importantly, our findings suggest that, although the flavor evolution depends on the mass ordering, contrarily to the conclusions reached in Ref.~\cite{Dasgupta:2017oko} by relying on a vanishing vacuum frequency, the final flavor configuration is not easily predictable. Given the highly non-linear nature of the system, it may be possible that for different angular distributions of neutrinos and antineutrinos, flavor conversions are enhanced in normal ordering instead than in inverted ordering, as shown in Fig.~\ref{Fig4}. This is another difference with respect to what found for  ``slow'' $\nu$--$\nu$ conversion where no flavor conversion is expected in normal ordering, see e.g.~\cite{Fogli:2007bk,Hannestad:2006nj}. 

\section{Impact of the vacuum mixing parameter: multi-energy configuration}
\label{sec5}
\begin{figure}
\includegraphics[width=0.49\textwidth]{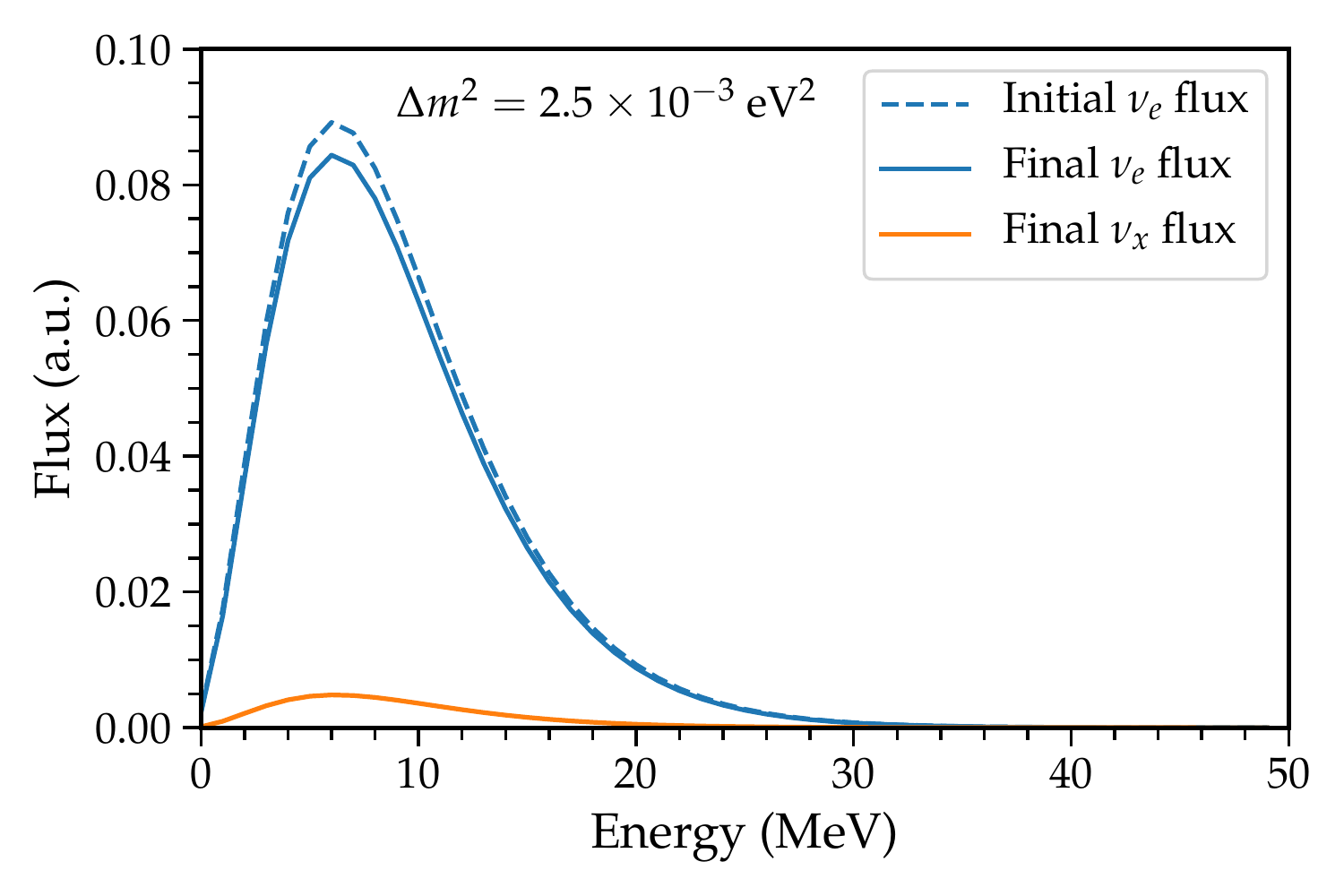}
\includegraphics[width=0.49\textwidth]{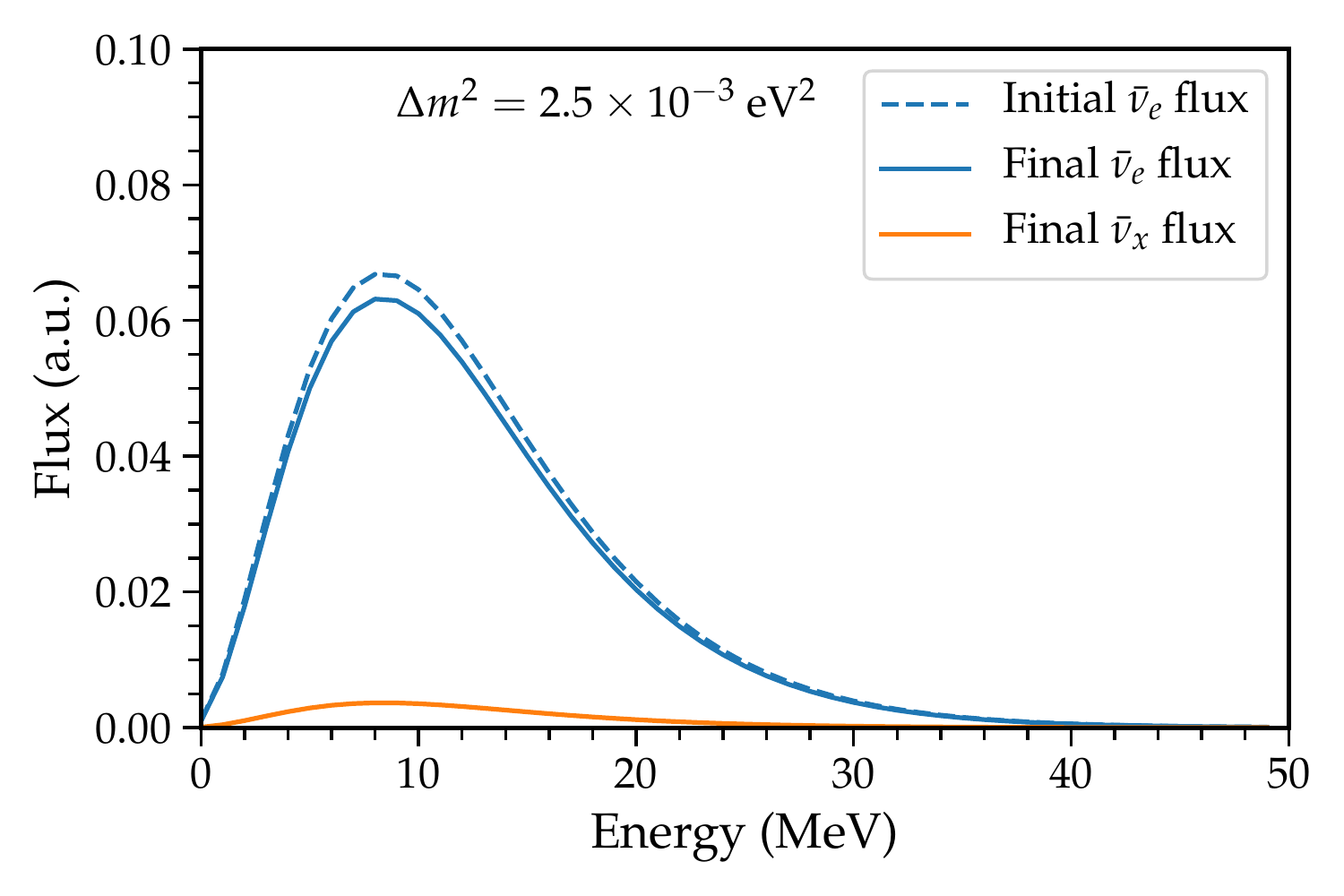}\\
\includegraphics[width=0.49\textwidth]{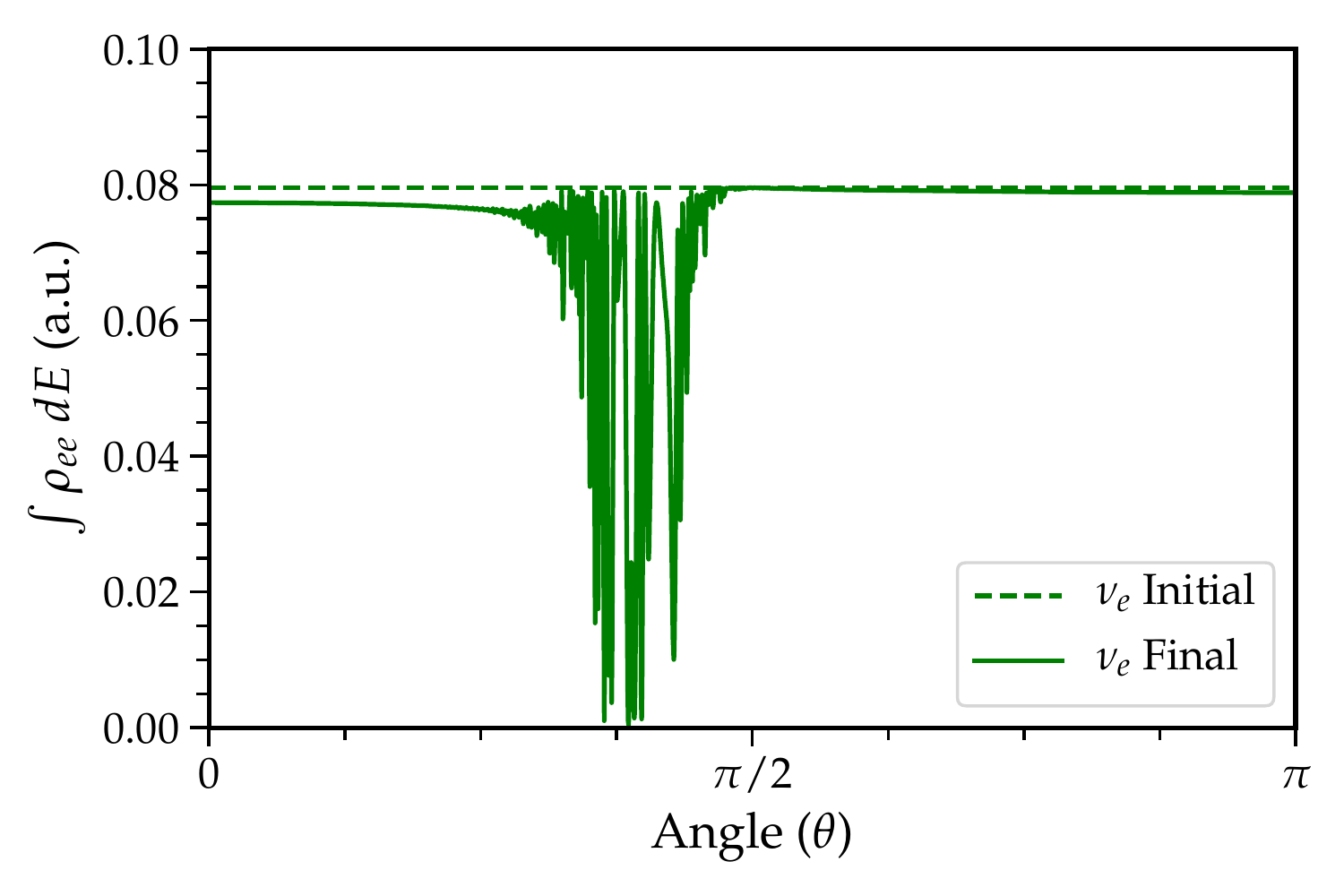}
\includegraphics[width=0.49\textwidth]{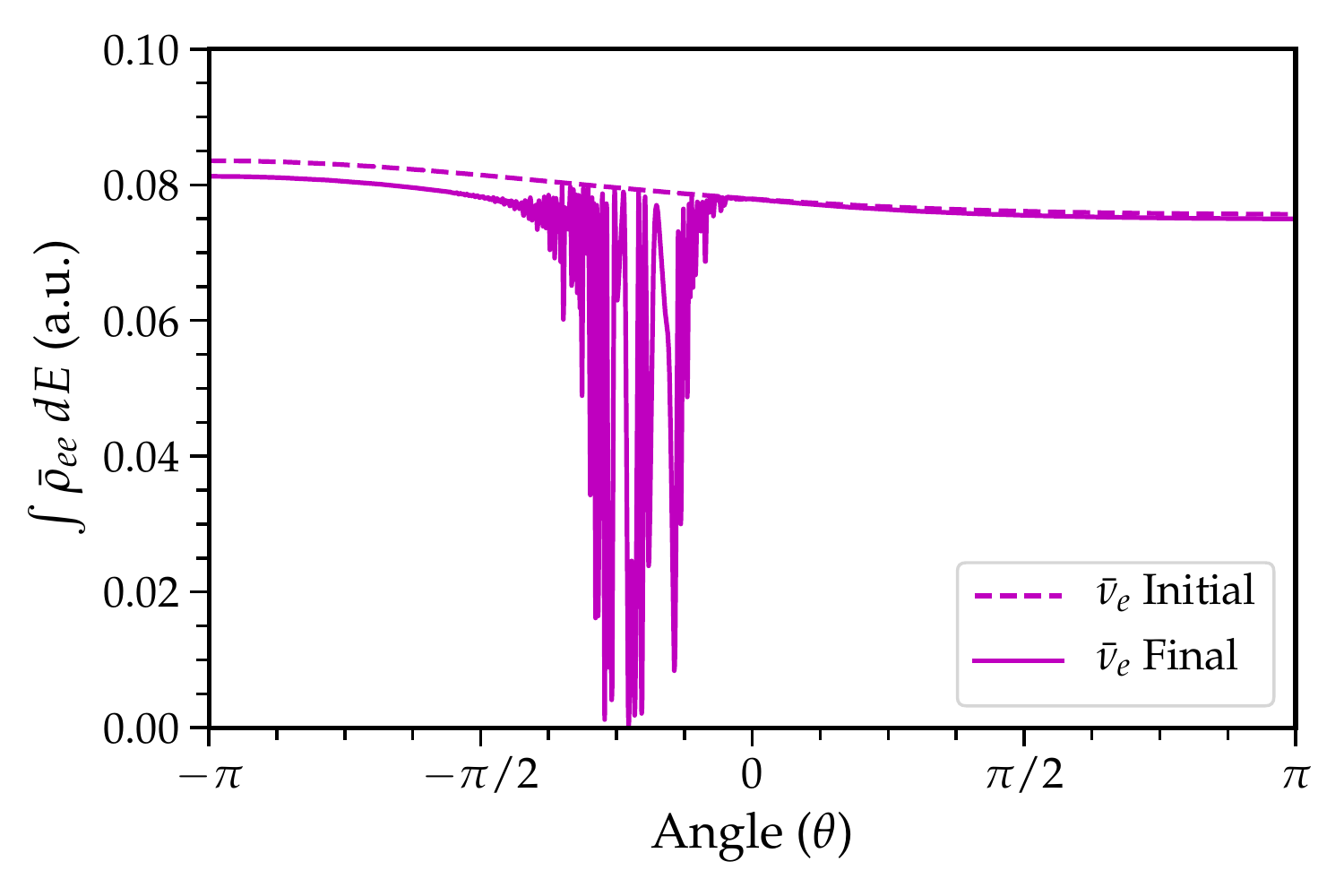}
\caption{{\it Top:} Spectral energy distributions for neutrinos (on the left) and antineutrinos (on the right) as  functions of the neutrino energy at $t=0$ (initial, dashed lines, see Eq.~\ref{eq:FD}) and $t=3\times 10^{-6}$~sec (final, solid lines), using $\Delta m^{2} = 2.5 \times 10^{-3}$ eV$^{2}$.
{\it Bottom:} Corresponding energy-averaged initial (dashed) and final (solid) angular distributions as functions of $\theta$.  Similarly to the single-energy case, fast flavor conversions in the multi-energy configuration are triggered by the ELN crossing present in the angular distributions.
}
\label{Fig5}
\end{figure}

In this section, we explore how the development of fast flavor conversions is further affected by  a multi-energy distribution for the (anti)neutrino ensemble. For the sake of simplicity, we consider an initial ensemble of $\nu_e$ and $\bar\nu_e$ (without non-electron flavor neutrinos) with the angular distributions introduced in Eq.~\ref{spec1} and with Fermi-Dirac energy distributions:
\begin{eqnarray}
\label{eq:FD}
n_{\nu_e, \bar\nu_e}(E) \propto \frac{E^{2}}{1+\exp(E/T_{\nu_e, \bar\nu_e})}\ ,
\end{eqnarray}
with  temperatures $T_{\bar\nu_e} = 4$~MeV and $T_{\nu_e} = 3$~MeV, see dashed lines in the top panels of Fig.~\ref{Fig5}.

The top panel of Fig.~\ref{Fig5} shows the evolution of the spectral energy distributions of neutrinos (on the left) and antineutrinos (on the right) for $\Delta m^2 = 2.5 \times 10^{-3}$~eV$^2$, the corresponding energy-averaged angular distributions are represented in the bottom panels. One can clearly see that the flavor conversions are triggered,  as it was found in Sec.~\ref{sec4}, by the ELN crossing present in the angular distributions. In turn, this induces modifications in the energy spectra. 

\begin{figure}
\includegraphics[width=0.49\textwidth]{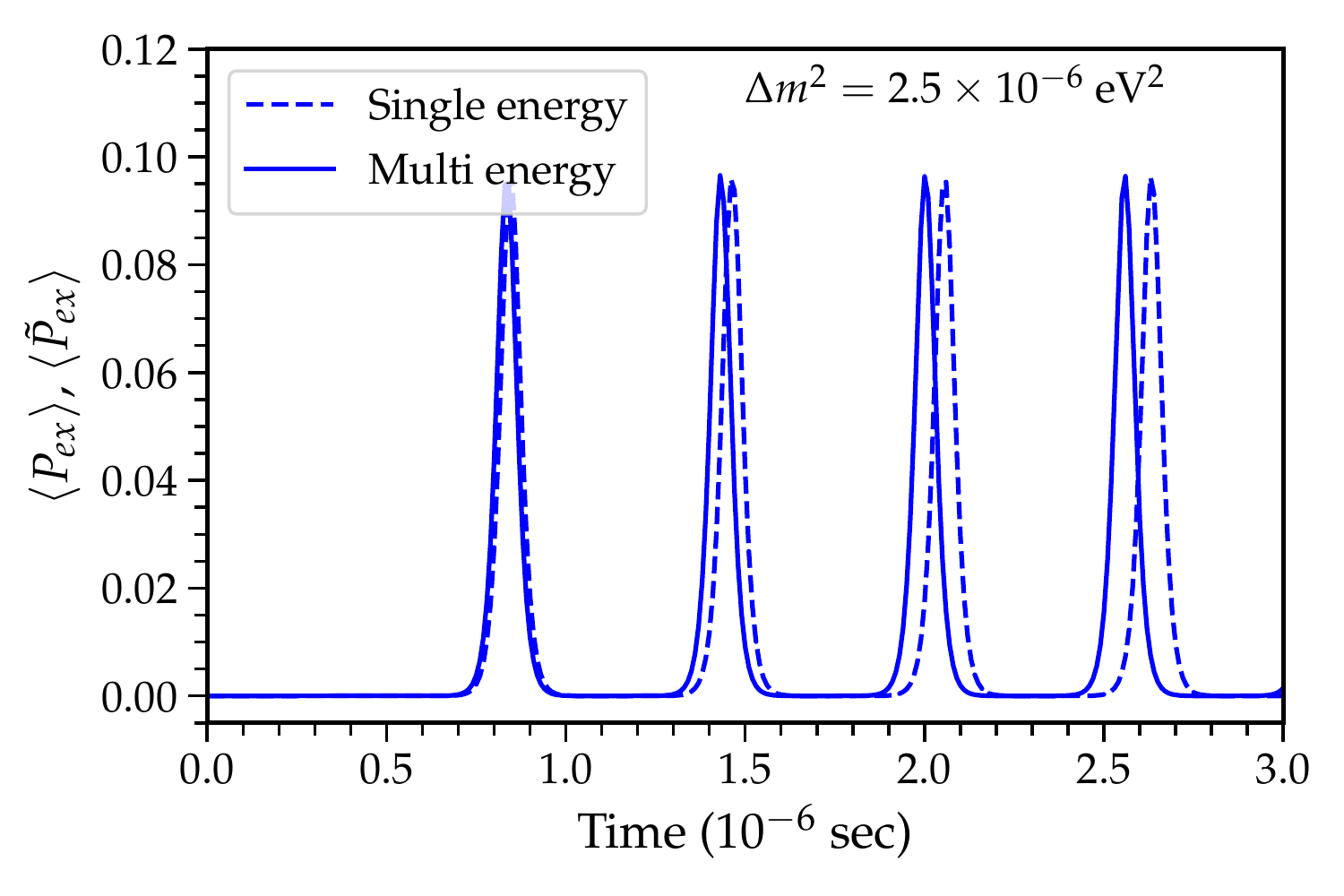}
\includegraphics[width=0.49\textwidth]{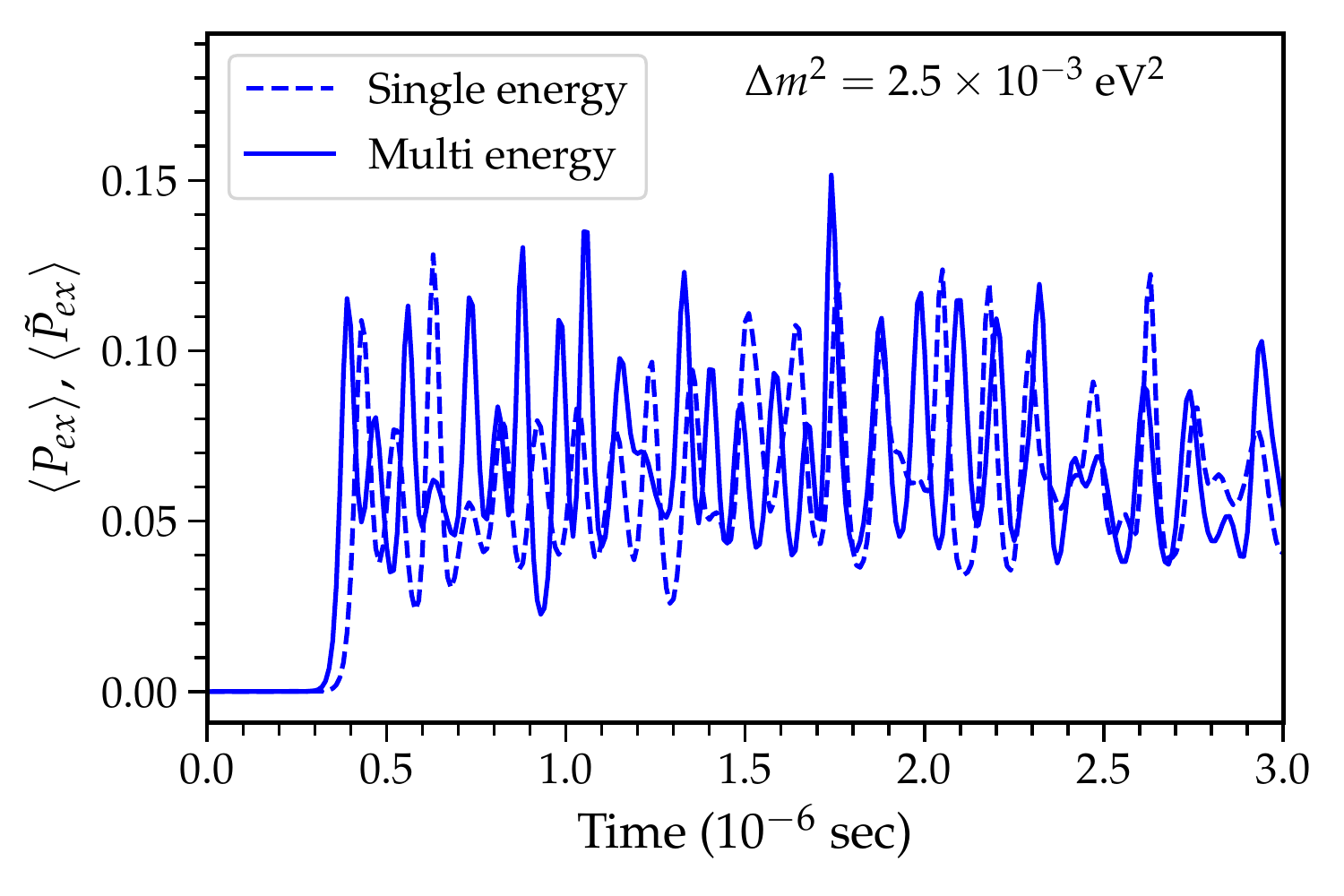}
\caption{Evolution of the angle-averaged (average-energy configuration) and angle- and energy-averaged transition probabilities as a function of time. The solid (dashed) lines represent   the multi-energy (average-energy) configuration.
The initial energy of $\nu_e$ and $\bar\nu_e$  in the average-energy case coincides with the average energy of the multi-energy setup; see text for details and Eq.~\ref{eq:FD}. The average-energy case qualitatively mimics the multi-energy configuration. The left (right) panel assumes $\Delta m^{2} = 2.5 \times 10^{-6}$ eV$^{2}$  ($\Delta m^{2} = 2.5 \times 10^{-3}$ eV$^{2}$);  the oscillation frequency increases as $\Delta m^2$ increases. The average-energy configurations give very similar results to the multi-energy ones, and no qualitative differences are observable. 
}
\label{Fig6}
\end{figure}

In  order to facilitate a comparison with the single-energy configuration discussed in Sec.~\ref{sec4} and better gauge the role played by the energy distribution in the final flavor configuration,  Fig.~\ref{Fig6} compares our multi-energy findings to the ones obtained for  neutrinos and antineutrinos having  one energy respectively, each correspondent to the average energy of the spectral distribution of Eq.~\ref{eq:FD}~\cite{Keil:2002in}, i.e., $E_{\nu_e} = 3.15 \times T_{\nu_e} = 9.45$~MeV and  $E_{\bar\nu_e} = 3.15 \times T_{\bar\nu_e} = 12.60$~MeV (average-energy configuration).  For both scenarios, we assume the angular distributions introduced in Eq.~\ref{spec1}.
The angle-averaged transition probability $\langle P_{ex} \rangle$ and its energy-integrated equivalent, $\langle \tilde P_{ex} \rangle$ (obtained by integrating over energy each term in Eq.~\ref{eq:Pex}), are shown in  Fig.~\ref{Fig6} for $\Delta m^2 = 2.5 \times 10^{-6}$~eV$^2$ (on the left) and $\Delta m^2 = 2.5 \times 10^{-3}$~eV$^2$ (on the right). 
One can see that, in the absence of crossings in the energy distributions, the average-energy configurations give results very similar to the multi-energy ones, with no qualitative differences. 
Unsurprisingly, when $\Delta m^2$ is small ($\Delta m^{2}=2.5\times 10^{-6}$ eV$^{2}$), we see a bipolar evolution of the neutrino flavor,  as discussed in Sec.~\ref{sec5}. The typical time-scale between two bipolar modes (time between two successive peaks of $\langle P_{ex} \rangle$) is comparable to the single-energy configuration, see Sec.~\ref{sec5}. Moreover, the oscillation frequency is strongly affected by  $\Delta m^2$ as the latter increases.

\section{Breaking of the  reflection symmetry}\label{sec6}
Throughout this paper, we have assumed azimuthal symmetry for the sake of simplicity.
The symmetry breaking  adds a new layer of complexity to the problem that goes beyond adding another term in the Hamiltonian and leads to the spatial symmetry breaking  of the system~\cite{Duan:2014gfa, Cirigliano:2017hmk,Shalgar:2019qwg}. 
In its most general form, the  $\nu$--$\nu$ term of the Hamiltonian   is 
\begin{eqnarray}
        H_{\nu\nu}(\theta,\phi) &=& \mu \int_{0}^{2\pi} \int_{0}^{\pi} \sin\theta^{\prime} d\theta^{\prime} d\phi^{\prime} \left[\rho(\theta^{\prime},\phi^{\prime})-\bar{\rho}(\theta^{\prime},\phi^{\prime})\right] \nonumber\\
&\times& \left[1-\cos\theta\cos\theta^{\prime}-\sin\theta\sin\theta^{\prime} \cos(\phi-\phi^{\prime})\right]\ .
        \label{Hnunu:full}
\end{eqnarray}
The additional term, with respect to Eq.~\ref{Hnunu:sym}, that depends on $\sin\theta$ may qualitatively change the neutrino flavor evolution and, thus, it should be included; however, at the same time, this additional term  makes the numerical simulations unfeasible. 

The effect of the additional $\sin\theta$  term in $H_{\nu\nu}$ can be mimicked without numerically solving the equations of motion by including the $\phi$ variable, but by  evolving the neutrino flavor in a two dimensional system with $\theta \in [0,2\pi]$ (instead than $\theta \in [0,\pi]$) and with the phase space integration factor  being $d\theta$ instead of $\sin\theta d\theta d\phi$.  
Then the resultant $\nu$--$\nu$ term of the Hamiltonian  is 
\begin{eqnarray}
H_{\nu\nu}(\theta) = \mu \int_{-\pi}^{\pi} d\theta^{\prime} [\rho(\theta^{\prime})-\bar{\rho}(\theta^{\prime})](1-\cos\theta\cos\theta^{\prime}-\sin\theta\sin\theta^{\prime})\ .
\label{Hnunu:asym}
\end{eqnarray}
This approximation, although restricted to a two-dimensional space, allows to investigate the possible effect of the term proportional to $\sin\theta$ in $H_{\nu\nu}$ on the flavor evolution without introducing the  azimuthal coordinate, when   the assumption of reflection symmetry ($\rho(\theta)=\rho(-\theta)$ and $\bar{\rho}(\theta)=\bar{\rho}(-\theta)$) is relaxed. 

In the numerical solution, we  include a small asymmetric perturbation of $\mathcal{O}(10^{-6})$ in the initial $\bar{\rho}_{ee}$. Moreover, we assume the  energy (angular) distribution  as in Eq.~\ref{eq:FD} (Eq.~\ref{spec1}).

\begin{figure}
\center
\includegraphics[width=0.49\textwidth]{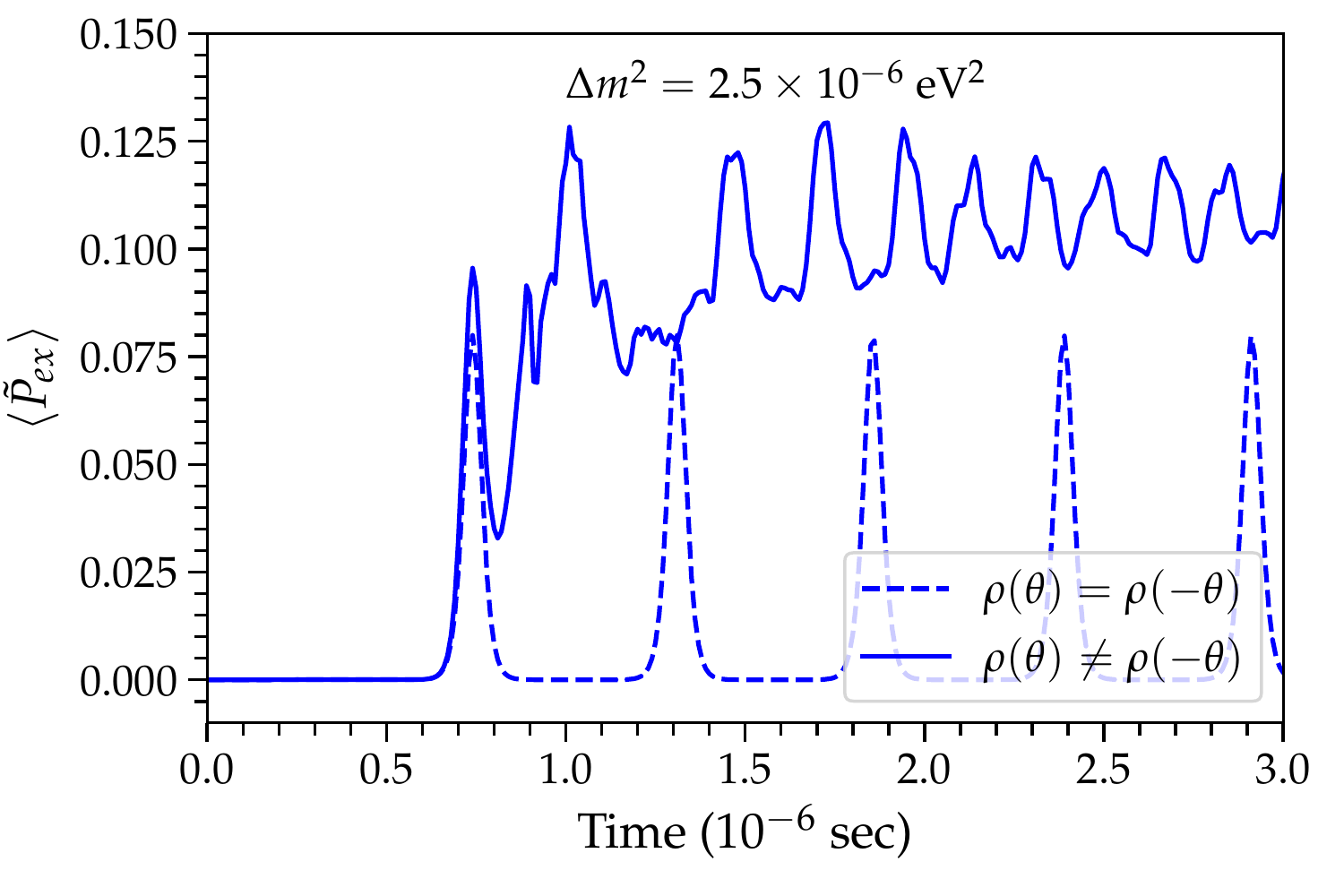}
\includegraphics[width=0.49\textwidth]{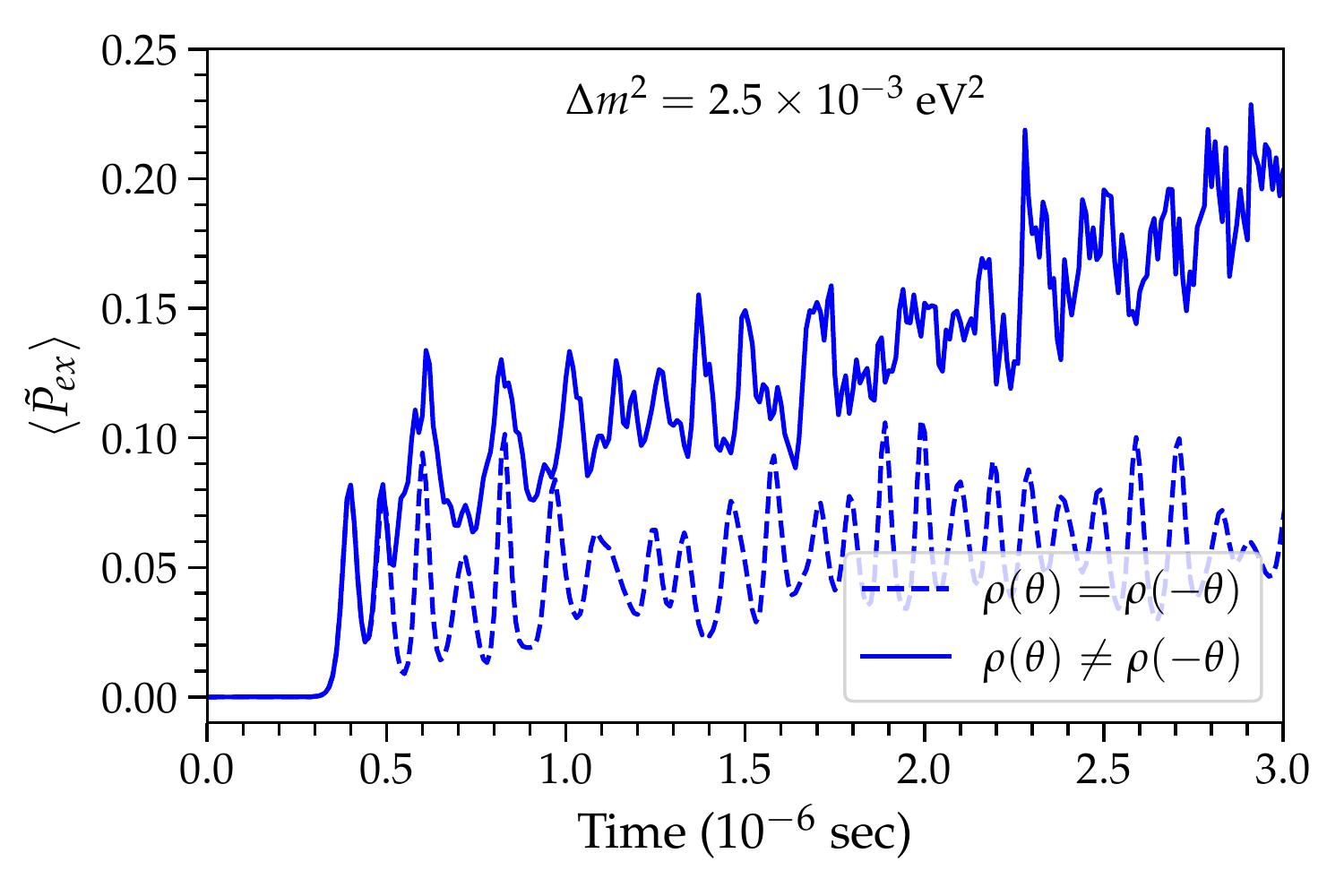}
\caption{Angle- and energy-averaged transition probability $\langle \tilde{P}_{ex} \rangle$ as  a function of time for $\Delta m^2 = 2.5 \times 10^{-6}$ eV$^{2}$ (on the left), $2.5 \times 10^{-6}$ eV$^{2}$ (on the right), when the reflection symmetry is broken (solid lines, $\rho(\theta)\neq \rho(-\theta)$) and when it is not broken (dashed lines, $\rho(\theta)=\rho(-\theta)$). The dependence of the transition probability on $\Delta m^2$ is still present, and the neutrino flavor evolution is no longer bipolar even for small $\Delta m^2$. 
}
\label{Fig7}
\end{figure}

Figure~\ref{Fig7} shows the energy and angle averaged transition probability $\langle \tilde{P}_{ex} \rangle$ as  a function of time for the cases where the reflection symmetry is imposed (dashed lines) and when it is broken (solid lines)  for $\Delta m^2 = 2.5 \times 10^{-6}$ eV$^{2}$ on the left and $\Delta m^2 = 2.5 \times 10^{-3}$ eV$^{2}$ on the right. One can see that there is still a  dependence on $\Delta m^2$.  However, as it was found for ``slow'' conversions~\cite{Raffelt:2013rqa}, the  inclusion of the asymmetric mode makes the system more unstable and  the neutrino flavor evolution is no longer bipolar, even for small  $\Delta m^2$. 

\section{Conclusions}
\label{sec7}
Pairwise flavor conversions of neutrinos are expected to occur in a dense neutrino gas and have been considered to be exclusively driven by the angular distributions of $\nu_e$ and $\bar\nu_e$.  
In particular, the existing literature and the massive employment of the linear stability analysis  have lent support to the fact that fast pairwise conversions can be triggered by the existence of crossings in the angular distributions of $\nu_e$ and $\bar\nu_e$ (ELN crossings) and that pairwise conversions can also occur in the absence of the neutrino mass splitting.

In this paper, we focus on the rich and enigmatic phenomenology of fast pairwise conversions of neutrinos and dispel some naive generalizations on their nature.  In particular, although the neutrino mass splitting has been traditionally neglected under the assumption that the vacuum frequency is a few orders of magnitude smaller than the $\nu$--$\nu$ potential, we show that fast pairwise conversions are actually  affected by the neutrino energy and are sensitive to the neutrino mass ordering, especially in the non-linear regime. Such a strong dependence on the neutrino vacuum frequency is surprising and counter-intuitive, it is discussed for the first time in this work for realistic values of the neutrino vacuum frequency in the linear and non-linear regime, and it highlights the highly non-linear nature of the dense neutrino gas.

Even if the energy-dependent linear stability analysis well reproduces the numerical solutions of the flavor evolution in the linear regime,  the vacuum term affects the non-linear regime and the onset of fast flavor conversions, inducing higher frequency modulations in the conversion probability as the squared mass difference $\Delta m^2$ increases (or, equivalently, the neutrino energy decreases). Interestingly, a regime very similar to the bipolar one~\cite{Hannestad:2006nj} is observed for $\Delta m^2 \rightarrow 0$~\cite{Dasgupta:2017oko,Johns:2019izj}, but any oscillation periodicity  disappears as $\Delta m^2$ increases approaching the atmospheric mass splitting or if  the reflection symmetry is broken. This makes the pendulum analogy, widely employed to model the physics of slow $\nu$--$\nu$ conversions, not suitable to grasp the basics of fast pairwise conversions. In addition, we find a strong dependence of the fast pairwise conversion transition probability on the mass ordering, differently from  what stated in Ref.~\cite{Dasgupta:2017oko} by relying on a vanishing vacuum frequency.

Contrary to what was found for  ``slow'' neutrino self-interactions, our work shows that the spectral energy distributions of neutrinos and antineutrinos do not affect the flavor evolution dramatically. In fact, a qualitatively similar behavior is found when the multi-energy configuration for neutrinos and antineutrinos is approximated by two energy modes, each corresponding to the average energy of the neutrino and antineutrino spectral distributions.
In the context of ``slow'' neutrino self-interactions, it was also shown that the relaxation of the  symmetries imposed on the system make the latter more unstable. Similarly, for fast pairwise conversions, we find that the inclusion of an asymmetric mode breaks any periodicity in the oscillation patterns and favors the growth of the instability.

In conclusion, our findings highlight the highly counter-intuitive phenomenology of fast pairwise conversions and further corroborate the need to move beyond simple semi-analytical treatments. Only in this  way will we be able to grasp the neutrino flavor evolution physics and finally gauge its relevance in astrophysical sources.

\acknowledgments
We are grateful to Georg Raffelt for helpful discussions. We acknowledge support from  the Villum Foundation (Project No.~13164), the Danmarks Frie Forskningsfonds (Project No.~8049-00038B), the Knud H\o jgaard Foundation, and the Deutsche Forschungsgemeinschaft through Sonderforschungbereich
SFB~1258 ``Neutrinos and Dark Matter in Astro- and
Particle Physics'' (NDM). 

%\begin{thebibliography}{99}
\bibliographystyle{JHEP}
\bibliography{omega3}

\providecommand{\href}[2]{#2}\begingroup\raggedright\begin{thebibliography}{10}

\bibitem{Mirizzi:2015eza}
A.~Mirizzi, I.~Tamborra, H.-T. Janka, N.~Saviano, K.~Scholberg, R.~Bollig
  et~al., \emph{{Supernova Neutrinos: Production, Oscillations and Detection}},
  \href{https://doi.org/10.1393/ncr/i2016-10120-8}{\emph{Riv. Nuovo Cim.}
  {\bfseries 39} (2016) 1--112},
  [\href{https://arxiv.org/abs/1508.00785}{{\ttfamily 1508.00785}}].

\bibitem{Horiuchi:2017sku}
S.~Horiuchi and J.~P. Kneller, \emph{{What can be learned from a future
  supernova neutrino detection?}},
  \href{https://doi.org/10.1088/1361-6471/aaa90a}{\emph{J. Phys. G} {\bfseries
  45} (2018) 043002}, [\href{https://arxiv.org/abs/1709.01515}{{\ttfamily
  1709.01515}}].

\bibitem{Chakraborty:2016yeg}
S.~Chakraborty, R.~Hansen, I.~Izaguirre and G.~G. Raffelt, \emph{{Collective
  neutrino flavor conversion: Recent developments}},
  \href{https://doi.org/10.1016/j.nuclphysb.2016.02.012}{\emph{Nucl. Phys. B}
  {\bfseries 908} (2016) 366--381},
  [\href{https://arxiv.org/abs/1602.02766}{{\ttfamily 1602.02766}}].

\bibitem{Duan:2010bg}
H.~Duan, G.~M. Fuller and Y.-Z. Qian, \emph{{Collective Neutrino
  Oscillations}},
  \href{https://doi.org/10.1146/annurev.nucl.012809.104524}{\emph{Ann. Rev.
  Nucl. Part. Sci.} {\bfseries 60} (2010) 569--594},
  [\href{https://arxiv.org/abs/1001.2799}{{\ttfamily 1001.2799}}].

\bibitem{Wu:2017qpc}
M.-R. Wu and I.~Tamborra, \emph{{Fast neutrino conversions: Ubiquitous in
  compact binary merger remnants}},
  \href{https://doi.org/10.1103/PhysRevD.95.103007}{\emph{Phys. Rev. D}
  {\bfseries 95} (2017) 103007},
  [\href{https://arxiv.org/abs/1701.06580}{{\ttfamily 1701.06580}}].

\bibitem{Johns:2016enc}
L.~Johns, M.~Mina, V.~Cirigliano, M.~W. Paris and G.~M. Fuller, \emph{{Neutrino
  flavor transformation in the lepton-asymmetric universe}},
  \href{https://doi.org/10.1103/PhysRevD.94.083505}{\emph{Phys. Rev. D}
  {\bfseries 94} (2016) 083505},
  [\href{https://arxiv.org/abs/1608.01336}{{\ttfamily 1608.01336}}].

\bibitem{Mikheev:1986gs}
S.~P. Mikheyev and A.~{\relax Yu}. Smirnov, \emph{{Resonance Amplification of
  Oscillations in Matter and Spectroscopy of Solar Neutrinos}}, {\emph{Sov. J.
  Nucl. Phys.} {\bfseries 42} (1985) 913--917}.

\bibitem{Wolfenstein:1977ue}
L.~Wolfenstein, \emph{{Neutrino Oscillations in Matter}},
  \href{https://doi.org/10.1103/PhysRevD.17.2369}{\emph{Phys. Rev.} {\bfseries
  D17} (1978) 2369--2374}.

\bibitem{Sigl:1992fn}
G.~Sigl and G.~G. Raffelt, \emph{{General kinetic description of relativistic
  mixed neutrinos}},
  \href{https://doi.org/10.1016/0550-3213(93)90175-O}{\emph{Nucl. Phys. B}
  {\bfseries 406} (1993) 423--451}.

\bibitem{Hannestad:2006nj}
S.~Hannestad, G.~G. Raffelt, G.~Sigl and Y.~Y. Wong, \emph{{Self-induced
  conversion in dense neutrino gases: Pendulum in flavour space}},
  \href{https://doi.org/10.1103/PhysRevD.74.105010}{\emph{Phys. Rev. D}
  {\bfseries 74} (2006) 105010},
  [\href{https://arxiv.org/abs/astro-ph/0608695}{{\ttfamily
  astro-ph/0608695}}].

\bibitem{Fogli:2007bk}
G.~L. Fogli, E.~Lisi, A.~Marrone and A.~Mirizzi, \emph{{Collective neutrino
  flavor transitions in supernovae and the role of trajectory averaging}},
  \href{https://doi.org/10.1088/1475-7516/2007/12/010}{\emph{JCAP} {\bfseries
  12} (2007) 010}, [\href{https://arxiv.org/abs/0707.1998}{{\ttfamily
  0707.1998}}].

\bibitem{Duan:2005cp}
H.~Duan, G.~M. Fuller and Y.-Z. Qian, \emph{{Collective neutrino flavor
  transformation in supernovae}},
  \href{https://doi.org/10.1103/PhysRevD.74.123004}{\emph{Phys. Rev.}
  {\bfseries D74} (2006) 123004},
  [\href{https://arxiv.org/abs/astro-ph/0511275}{{\ttfamily
  astro-ph/0511275}}].

\bibitem{Duan:2006jv}
H.~Duan, G.~M. Fuller, J.~Carlson and Y.-Z. Qian, \emph{{Coherent Development
  of Neutrino Flavor in the Supernova Environment}},
  \href{https://doi.org/10.1103/PhysRevLett.97.241101}{\emph{Phys. Rev. Lett.}
  {\bfseries 97} (2006) 241101},
  [\href{https://arxiv.org/abs/astro-ph/0608050}{{\ttfamily
  astro-ph/0608050}}].

\bibitem{Duan:2006an}
H.~Duan, G.~M. Fuller, J.~Carlson and Y.-Z. Qian, \emph{{Simulation of Coherent
  Non-Linear Neutrino Flavor Transformation in the Supernova Environment. 1.
  Correlated Neutrino Trajectories}},
  \href{https://doi.org/10.1103/PhysRevD.74.105014}{\emph{Phys. Rev.}
  {\bfseries D74} (2006) 105014},
  [\href{https://arxiv.org/abs/astro-ph/0606616}{{\ttfamily
  astro-ph/0606616}}].

\bibitem{Banerjee:2011fj}
A.~Banerjee, A.~Dighe and G.~G. Raffelt, \emph{{Linearized flavor-stability
  analysis of dense neutrino streams}},
  \href{https://doi.org/10.1103/PhysRevD.84.053013}{\emph{Phys. Rev. D}
  {\bfseries 84} (2011) 053013},
  [\href{https://arxiv.org/abs/1107.2308}{{\ttfamily 1107.2308}}].

\bibitem{Sawyer:2015dsa}
R.~F. Sawyer, \emph{{Neutrino cloud instabilities just above the neutrino
  sphere of a supernova}},
  \href{https://doi.org/10.1103/PhysRevLett.116.081101}{\emph{Phys. Rev. Lett.}
  {\bfseries 116} (2016) 081101},
  [\href{https://arxiv.org/abs/1509.03323}{{\ttfamily 1509.03323}}].

\bibitem{Sawyer:2005jk}
R.~F. Sawyer, \emph{{Speed-up of neutrino transformations in a supernova
  environment}}, \href{https://doi.org/10.1103/PhysRevD.72.045003}{\emph{Phys.
  Rev. D} {\bfseries 72} (2005) 045003},
  [\href{https://arxiv.org/abs/hep-ph/0503013}{{\ttfamily hep-ph/0503013}}].

\bibitem{Sawyer:2008zs}
R.~F. Sawyer, \emph{{The multi-angle instability in dense neutrino systems}},
  \href{https://doi.org/10.1103/PhysRevD.79.105003}{\emph{Phys. Rev.}
  {\bfseries D79} (2009) 105003},
  [\href{https://arxiv.org/abs/0803.4319}{{\ttfamily 0803.4319}}].

\bibitem{Chakraborty:2016lct}
S.~Chakraborty, R.~S. Hansen, I.~Izaguirre and G.~Raffelt, \emph{{Self-induced
  neutrino flavor conversion without flavor mixing}},
  \href{https://doi.org/10.1088/1475-7516/2016/03/042}{\emph{JCAP} {\bfseries
  03} (2016) 042}, [\href{https://arxiv.org/abs/1602.00698}{{\ttfamily
  1602.00698}}].

\bibitem{Dasgupta:2016dbv}
B.~Dasgupta, A.~Mirizzi and M.~Sen, \emph{{Fast neutrino flavor conversions
  near the supernova core with realistic flavor-dependent angular
  distributions}},
  \href{https://doi.org/10.1088/1475-7516/2017/02/019}{\emph{JCAP} {\bfseries
  1702} (2017) 019}, [\href{https://arxiv.org/abs/1609.00528}{{\ttfamily
  1609.00528}}].

\bibitem{Izaguirre:2016gsx}
I.~Izaguirre, G.~G. Raffelt and I.~Tamborra, \emph{{Fast Pairwise Conversion of
  Supernova Neutrinos: A Dispersion-Relation Approach}},
  \href{https://doi.org/10.1103/PhysRevLett.118.021101}{\emph{Phys. Rev. Lett.}
  {\bfseries 118} (2017) 021101},
  [\href{https://arxiv.org/abs/1610.01612}{{\ttfamily 1610.01612}}].

\bibitem{Tamborra:2017ubu}
I.~Tamborra, L.~H{\"u}depohl, G.~G. Raffelt and H.-T. Janka,
  \emph{{Flavor-dependent neutrino angular distribution in core-collapse
  supernovae}},
  \href{https://doi.org/10.3847/1538-4357/aa6a18}{\emph{Astrophys. J.}
  {\bfseries 839} (2017) 132},
  [\href{https://arxiv.org/abs/1702.00060}{{\ttfamily 1702.00060}}].

\bibitem{Shalgar:2019kzy}
S.~Shalgar and I.~Tamborra, \emph{{On the Occurrence of Crossings Between the
  Angular Distributions of Electron Neutrinos and Antineutrinos in the
  Supernova Core}},
  \href{https://doi.org/10.3847/1538-4357/ab38ba}{\emph{Astrophys. J.}
  {\bfseries 883} (2019) 80},
  [\href{https://arxiv.org/abs/1904.07236}{{\ttfamily 1904.07236}}].

\bibitem{Abbar:2019zoq}
S.~Abbar, H.~Duan, K.~Sumiyoshi, T.~Takiwaki and M.~C. Volpe, \emph{{Fast
  Neutrino Flavor Conversion Modes in Multidimensional Core-collapse Supernova
  Models: the Role of the Asymmetric Neutrino Distributions}},
  \href{https://doi.org/10.1103/PhysRevD.101.043016}{\emph{Phys. Rev. D}
  {\bfseries 101} (2020) 043016},
  [\href{https://arxiv.org/abs/1911.01983}{{\ttfamily 1911.01983}}].

\bibitem{Abbar:2018shq}
S.~Abbar, H.~Duan, K.~Sumiyoshi, T.~Takiwaki and M.~C. Volpe, \emph{{On the
  occurrence of fast neutrino flavor conversions in multidimensional supernova
  models}}, \href{https://doi.org/10.1103/PhysRevD.100.043004}{\emph{Phys.
  Rev.} {\bfseries D100} (2019) 043004},
  [\href{https://arxiv.org/abs/1812.06883}{{\ttfamily 1812.06883}}].

\bibitem{DelfanAzari:2019tez}
M.~Delfan~Azari, S.~Yamada, T.~Morinaga, H.~Nagakura, S.~Furusawa, A.~Harada
  et~al., \emph{{Fast collective neutrino oscillations inside the neutrino
  sphere in core-collapse supernovae}},
  \href{https://doi.org/10.1103/PhysRevD.101.023018}{\emph{Phys. Rev. D}
  {\bfseries 101} (2020) 023018},
  [\href{https://arxiv.org/abs/1910.06176}{{\ttfamily 1910.06176}}].

\bibitem{Nagakura:2019sig}
H.~{Nagakura}, T.~{Morinaga}, C.~{Kato} and S.~{Yamada}, \emph{{Fast-pairwise
  Collective Neutrino Oscillations Associated with Asymmetric Neutrino
  Emissions in Core-collapse Supernovae}},
  \href{https://doi.org/10.3847/1538-4357/ab4cf2}{\emph{Astrophys. J.}
  {\bfseries 886} (Dec., 2019) 139},
  [\href{https://arxiv.org/abs/1910.04288}{{\ttfamily 1910.04288}}].

\bibitem{Morinaga:2019wsv}
T.~Morinaga, H.~Nagakura, C.~Kato and S.~Yamada, \emph{{Fast neutrino-flavor
  conversion in the preshock region of core-collapse supernovae}},
  \href{https://doi.org/10.1103/PhysRevResearch.2.012046}{\emph{Phys. Rev.
  Res.} {\bfseries 2} (2020) 012046},
  [\href{https://arxiv.org/abs/1909.13131}{{\ttfamily 1909.13131}}].

\bibitem{Wu:2017drk}
M.-R. Wu, I.~Tamborra, O.~Just and H.-T. Janka, \emph{{Imprints of
  neutrino-pair flavor conversions on nucleosynthesis in ejecta from
  neutron-star merger remnants}},
  \href{https://doi.org/10.1103/PhysRevD.96.123015}{\emph{Phys. Rev. D}
  {\bfseries 96} (2017) 123015},
  [\href{https://arxiv.org/abs/1711.00477}{{\ttfamily 1711.00477}}].

\bibitem{Xiong:2020ntn}
Z.~Xiong, A.~Sieverding, M.~Sen and Y.-Z. Qian, \emph{{Potential Impact of Fast
  Flavor Oscillations on Neutrino-driven Winds and Their Nucleosynthesis}},
  \href{https://arxiv.org/abs/2006.11414}{{\ttfamily 2006.11414}}.

\bibitem{Bhattacharyya:2020dhu}
S.~Bhattacharyya and B.~Dasgupta, \emph{{Late-time behavior of fast neutrino
  oscillations}},
  \href{https://doi.org/10.1103/PhysRevD.102.063018}{\emph{Phys. Rev. D}
  {\bfseries 102} (2020) 063018},
  [\href{https://arxiv.org/abs/2005.00459}{{\ttfamily 2005.00459}}].

\bibitem{Abbar:2017pkh}
S.~Abbar and H.~Duan, \emph{{Fast neutrino flavor conversion: roles of dense
  matter and spectrum crossing}},
  \href{https://doi.org/10.1103/PhysRevD.98.043014}{\emph{Phys. Rev.}
  {\bfseries D98} (2018) 043014},
  [\href{https://arxiv.org/abs/1712.07013}{{\ttfamily 1712.07013}}].

\bibitem{Abbar:2020fcl}
S.~Abbar, \emph{{Searching for Fast Neutrino Flavor Conversion Modes in
  Core-collapse Supernova Simulations}},
  \href{https://doi.org/10.1088/1475-7516/2020/05/027}{\emph{JCAP} {\bfseries
  05} (2020) 027}, [\href{https://arxiv.org/abs/2003.00969}{{\ttfamily
  2003.00969}}].

\bibitem{Dasgupta:2018ulw}
B.~Dasgupta, A.~Mirizzi and M.~Sen, \emph{{Simple method of diagnosing fast
  flavor conversions of supernova neutrinos}},
  \href{https://doi.org/10.1103/PhysRevD.98.103001}{\emph{Phys. Rev. D}
  {\bfseries 98} (2018) 103001},
  [\href{https://arxiv.org/abs/1807.03322}{{\ttfamily 1807.03322}}].

\bibitem{Johns:2019izj}
L.~Johns, H.~Nagakura, G.~M. Fuller and A.~Burrows, \emph{{Neutrino
  oscillations in supernovae: angular moments and fast instabilities}},
  \href{https://doi.org/10.1103/PhysRevD.101.043009}{\emph{Phys. Rev. D}
  {\bfseries 101} (2020) 043009},
  [\href{https://arxiv.org/abs/1910.05682}{{\ttfamily 1910.05682}}].

\bibitem{Martin:2019gxb}
J.~D. Martin, C.~Yi and H.~Duan, \emph{{Dynamic fast flavor oscillation waves
  in dense neutrino gases}},
  \href{https://doi.org/10.1016/j.physletb.2019.135088}{\emph{Phys. Lett.}
  {\bfseries B800} (2020) 135088},
  [\href{https://arxiv.org/abs/1909.05225}{{\ttfamily 1909.05225}}].

\bibitem{Yi:2019hrp}
C.~Yi, L.~Ma, J.~D. Martin and H.~Duan, \emph{{Dispersion relation of the fast
  neutrino oscillation wave}},
  \href{https://doi.org/10.1103/PhysRevD.99.063005}{\emph{Phys. Rev. D}
  {\bfseries 99} (2019) 063005},
  [\href{https://arxiv.org/abs/1901.01546}{{\ttfamily 1901.01546}}].

\bibitem{Capozzi:2017gqd}
F.~Capozzi, B.~Dasgupta, E.~Lisi, A.~Marrone and A.~Mirizzi, \emph{{Fast flavor
  conversions of supernova neutrinos: Classifying instabilities via dispersion
  relations}}, \href{https://doi.org/10.1103/PhysRevD.96.043016}{\emph{Phys.
  Rev. D} {\bfseries 96} (2017) 043016},
  [\href{https://arxiv.org/abs/1706.03360}{{\ttfamily 1706.03360}}].

\bibitem{Shalgar:2019qwg}
S.~Shalgar, I.~Padilla-Gay and I.~Tamborra, \emph{{Neutrino propagation hinders
  fast pairwise flavor conversions}},
  \href{https://doi.org/10.1088/1475-7516/2020/06/048}{\emph{JCAP} {\bfseries
  06} (2020) 048}, [\href{https://arxiv.org/abs/1911.09110}{{\ttfamily
  1911.09110}}].

\bibitem{Dasgupta:2017oko}
B.~Dasgupta and M.~Sen, \emph{{Fast Neutrino Flavor Conversion as Oscillations
  in a Quartic Potential}},
  \href{https://doi.org/10.1103/PhysRevD.97.023017}{\emph{Phys. Rev. D}
  {\bfseries 97} (2018) 023017},
  [\href{https://arxiv.org/abs/1709.08671}{{\ttfamily 1709.08671}}].

\bibitem{EstebanPretel:2008ni}
A.~Esteban-Pretel, A.~Mirizzi, S.~Pastor, R.~Tom{\`a}s, G.~Raffelt, P.~Serpico
  et~al., \emph{{Role of dense matter in collective supernova neutrino
  transformations}},
  \href{https://doi.org/10.1103/PhysRevD.78.085012}{\emph{Phys. Rev. D}
  {\bfseries 78} (2008) 085012},
  [\href{https://arxiv.org/abs/0807.0659}{{\ttfamily 0807.0659}}].

\bibitem{Airen:2018nvp}
S.~Airen, F.~Capozzi, S.~Chakraborty, B.~Dasgupta, G.~Raffelt and T.~Stirner,
  \emph{{Normal-mode Analysis for Collective Neutrino Oscillations}},
  \href{https://doi.org/10.1088/1475-7516/2018/12/019}{\emph{JCAP} {\bfseries
  12} (2018) 019}, [\href{https://arxiv.org/abs/1809.09137}{{\ttfamily
  1809.09137}}].

\bibitem{Abbar:2018beu}
S.~Abbar and M.~C. Volpe, \emph{{On Fast Neutrino Flavor Conversion Modes in
  the Nonlinear Regime}},
  \href{https://doi.org/10.1016/j.physletb.2019.02.002}{\emph{Phys. Lett. B}
  {\bfseries 790} (2019) 545--550},
  [\href{https://arxiv.org/abs/1811.04215}{{\ttfamily 1811.04215}}].

\bibitem{Keil:2002in}
M.~T. Keil, G.~G. Raffelt and H.-T. Janka, \emph{{Monte Carlo study of
  supernova neutrino spectra formation}},
  \href{https://doi.org/10.1086/375130}{\emph{Astrophys. J.} {\bfseries 590}
  (2003) 971--991}, [\href{https://arxiv.org/abs/astro-ph/0208035}{{\ttfamily
  astro-ph/0208035}}].

\bibitem{Duan:2014gfa}
H.~Duan and S.~Shalgar, \emph{{Flavor instabilities in the neutrino line
  model}}, \href{https://doi.org/10.1016/j.physletb.2015.05.057}{\emph{Phys.
  Lett.} {\bfseries B747} (2015) 139--143},
  [\href{https://arxiv.org/abs/1412.7097}{{\ttfamily 1412.7097}}].

\bibitem{Cirigliano:2017hmk}
V.~Cirigliano, M.~W. Paris and S.~Shalgar, \emph{{Effect of collisions on
  neutrino flavor inhomogeneity in a dense neutrino gas}},
  \href{https://doi.org/10.1016/j.physletb.2017.09.039}{\emph{Phys. Lett. B}
  {\bfseries 774} (2017) 258--267},
  [\href{https://arxiv.org/abs/1706.07052}{{\ttfamily 1706.07052}}].

\bibitem{Raffelt:2013rqa}
G.~G. Raffelt, S.~Sarikas and D.~de~Sousa~Seixas, \emph{{Axial Symmetry
  Breaking in Self-Induced Flavor Conversion of Supernova Neutrino Fluxes}},
  \href{https://doi.org/10.1103/PhysRevLett.111.091101}{\emph{Phys. Rev. Lett.}
  {\bfseries 111} (2013) 091101},
  [\href{https://arxiv.org/abs/1305.7140}{{\ttfamily 1305.7140}}].

\end{thebibliography}\endgroup
%\end{thebibliography}
\end{document}